\documentclass[twoside,12pt]{article}
\usepackage[]{lineno}
\usepackage{epsfig}
\begin{document}
\vspace{0.5cm}
\begin{center}
\begin{Large}
\bf{Future Deep Inelastic Scattering with the LHeC} \\
\end{Large}
\vspace{0.5cm}
 Max Klein (University of Liverpool) \\
\vspace{0.2cm}
Contribution to a Book dedicated to the Memory of Guido Altarelli,  January 21, 2018 \\
\abstract{
\noindent
For nearly a decade, Guido Altarelli accompanied the Large Hadron electron Collider project, as invited speaker, referee and member of the International  Advisory Committee. This text summarises the status and prospects of the development of the LHeC, with admiration for a one-time scientist and singular leader whom I met first  nearly 40 years ago under the sun shining for the ``Herceg Novi School" in Kupari, where we both lectured about the beautiful science of Deep Inelastic Scattering and enjoyed life under a yellow moon. 
}
\end{center}
%
\section{Introduction}
The time is now $50$ years after the birth of deep inelastic scattering (DIS)
with the discovery of partons~\cite{Bloom:1969kc,Breidenbach:1969kd}
and $40$ years after the paper of Altarelli and Parisi, cited over six thousand times,
on ``Asymptotic Freedom in Parton Language"~\cite{Altarelli:1977zs}.
We are 20 years after the approval of the huge LHC detectors, ATLAS and CMS. These have
now taken about $100$\,fb$^{-1}$ of data each. Adding the publications of all LHC
collaborations, nearly two thousand papers have appeared, based on the superb performance
of the Large Hadron Collider. The Higgs boson was discovered five years ago. 
About this most impressive harvest Guido Altarelli
had noted, {\it{we expected complexity but we found a maximum of
simplicity~\cite{Altarelli:2014xxa}}}, commenting
not on the hugely complex experiments nor ingenuitive, novel analysis techniques 
but refering to the 
surprising absence of new physics, besides the exciting observation of the Higgs boson. No evidence was observed so far in support for extra dimensions, symmetries, 
particles nor a grand unified theory. 
Guido thought about this deeply: 
{\it{..it is true that the SM theory is renormalizable
and completely finite and predictive. If you forget the required miraculous fine
tuning you are not punished, you find no catastrophe! .. The possibility that the
Standard Model holds well beyond the electroweak scale must now be seriously considered..
 We are
experiencing a very puzzling situation but, to some extent, this is good because
big steps forward in fundamental physics have often originated from paradoxes.
We highly hope that the continuation of the LHC experiments will bring new
light on these problems...}}~\cite{Altarelli:2014xxa}.

The LHC is now projected to operate for two decades hence, with a major
interrupt for accelerator and detector upgrades in between.
A new horizon for these explorations would be opened with an additional $ep/eA$ experiment, i.e. with electron-proton (and electron-ion) collisions  concurrent to
the default LHC operation in the thirties.
The Large Hadron electron Collider (LHeC) was principally designed in an extended 
concept report (CDR)~\cite{Abelleira2012}. This was published  in June 2012, a few weeks before the Higgs boson discovery was announced. It 
considers the addition to the LHC of a 60\,GeV energy electron beam accelerator, arranged tangential to
the main ring of $27$\,km circumference. This configuration 
is thought to form a novel $ep$ and $eA$ experiment which would 
be able to collect O($1$)\,ab$^{-1}$ of integrated luminosity, exceeding the HERA
result by a factor of thousand and its kinematic range by nearly $20$. Its salient feature will be the synchronous operation of $ep$ with $pp$ such that no principal loss of LHC luminosity shall occur.
The LHeC would enable electron-hadron energy frontier physics explorations at 
about $1$\,TeV cms energy,  which is four times higher than future Higgs facilities, 
such as in China (CepC) or Japan (ILC), 
may achieve in electron-positron collisions. 
The question such a programme of deep inelastic electron-hadron scattering at the LHC
has to answer consists of how that may lead to new insight into the SM and beyond, on its own and in complementing
the $pp$ and heavy ion physics pursued at the Large Hadron Collider. This paper
has been written in memory of Guido Altarelli who advised the LHeC development
in his unforgettable manner as a member of the International Advisory Committee, a referee to the CDR and with many illuminating presentations and discussions. Together with Lev Nikolaevitch Lipatov, whom we lost so early also, the LHeC had two monumental theoretical physicists on its side, their inspiration, far reaching theoretical insight and high demand for quality.

The article presents brief accounts of the achievements of HERA (Sect.\,2) as well as lessons
from the first years of LHC physics (Sect.\,3). There follows a description of the LHeC 
accelerator design including
an introduction to its planned testfacility~\cite{Angal-Kalinin:2017iup} 
and a sketch of its detector concept~(Sect.\,4).
The LHeC would be the fifth major experiment 
at the LHC and it has five major themes which are illustrated in 
five subsequent sections:
\begin{itemize}
\item
Based on the unique hadron beams of the LHC and employing a point-like
probe,
it would represent the world's cleanest, high resolution
microscope for exploring the substructure of and dynamics inside matter
(Sect.\,5); 
\item
With concurrent $ep$ and $pp$ operation, the LHeC would transform the
LHC into a 3-beam, twin-collider, energy frontier accelerator facility. 
Through ultra-precise strong and electroweak interaction measurements,
the $ep$ results would make the HL LHC facility  a much more powerful
search and measurement laboratory than current expectations based on $pp$
only could possibly entail (Sect.\,6); 
\item
The clean DIS final state in neutral (NC) and charged currents (CC) enables a
high precision Higgs physics programme with the LHeC if it reaches 
O($10^{34}$) cm$^{-2}$s$^{-1}$ luminosity.
 The joint $pp/ep$ LHC facility can become a Higgs factory of unprecedented
impact. It would combine high, per cent precision for abundant Higgs decay channels
 in $ep$ with improved precision on rare channels,  high $pp$ luminosity and
the resolution of QCD uncertainties through $ep$. This  transforms the 
view on the Higgs physics potential at the LHC (Sect.\,7);  
\item
As  a new, unique, luminous TeV scale
collider, the LHeC has an outstanding opportunity to discover new
physics, such as in the exotic Higgs, dark matter, heavy neutrino and QCD
areas, which is actively being studied and will be illustrated in Sect.\,8; 
\item
The LHeC leads into the region of high parton densities at small Bjorken $x$.
It extends the kinematic range in lepton-nucleus scattering 
by nearly four orders of magnitude. It so is expected to  transform 
nuclear particle physics completely, by establishing a QCD base for Quark Gluon  Plasma (QGP)
phenomena and resolving the hitherto hidden parton dynamics and structure of nuclei (Sect.\,9).
\end{itemize}

Guido Altarelli delivered his first lecture to the LHeC community  at the first 
CERN-ECFA workshop in 2008~\cite{guido08}, in which he posed questions to work on. 
In June 2015 he delivered the last scientific presentation of his lifetime, 
actually to the annual workshop on 
the LHeC, when all of us had no clue about his health situation. There followed a
session of the International Advisory Committee for the LHeC, chaired by 
CERN DG emeritus, Herwig Schopper, in which Guido had asked: {\it{The physics
case of the LHeC is essentially made, can we proceed and build it?}} He knew
there would not be a straight answer possible then as we were all aware of things
still to be done. This article is an attempt
to describe the background of this statement and to sketch how indeed we 
can build a TeV scale  $ep/eA$ experiment, for the HL LHC lifetime and beyond into 
the HE LHC, should our community find the courage and means for supporting it. Some 
concluding remarks are made in Sect.\,10.

Guido Altarelli  often  raised the questions about the value and 
determination of the strong coupling constant $\alpha_s$ and also, 
 since his youth~\cite{Altarelli:1978tq}, why the 
longitudinal structure function $F_L$ was so hard to obtain. Having spent 
many years of  my life to measure both, I have added an Appendix, written 
from the  perspective of HERA and the LHeC, to this
otherwise more general overview.  

\section{Lessons from HERA}
It has become almost a habit to present $pp$ as the discovery machines and $e^+e^-$
as the clean machines, and to forget to mention $ep$. Be it intention or sloppiness, this has
been so often repeated that one may get tired
in raising the concern that such a view is indeed unjust with respect to the roles
of TeVatron, LEP and HERA, or the role of HERA for the LHC physics, and that it 
will hinder the prosperity of our science if its future is unwarily narrowed.
 Guido Altarelli  knew that a $2 \rightarrow 2$ scattering of $e$ and $p$
could be viewed as $ee$, $ep$ and $pp$, just as it had three Mandelstam variables
($s$, $t$ and $u$), and all three reactions had their merits and complementary roles because
the electron is a point-like lepton and the proton a composite hadron. The
principal recognition of the LHeC development is that $pp$ and $ep$ are a
symbiotic unit of different entities, which together would reach out much further
than alone. A combined LH(e)C facility, for example, would 
challenge a possible future $e^+e^-$ collider as $ep/pp$ combined with 
$e^+e^-$ would unravel the Higgs properties and the possible Higgs link to new physics
particularly thoroughly. The HERA  $ep$ collider could have discovered the Higgs
boson had it had a thousand times higher luminosity, which is the luminosity goal now. 
The LHeC will surpass its physics by an enormous extent, but it rests on HERA's foundations.

The ``Hochenergie Ringanlage" (HERA) was the first electron-proton collider
ever built. Its proposal  was endorsed in 1984.  HERA was operated between 1992 and 2007 
with colliding electron (also positron) and proton
beams,  of energy, for most of the time, of
 $E_e=26.7$\,GeV and $E_p=920$\,GeV.
The cms energy was $\sqrt{s}=2 \sqrt{E_eE_p}=319$\,GeV with which
HERA was suited to explore the physics at the Fermi scale, 
set by the vacuum expectation value  of the Higgs field,  $v=(G_F\sqrt{2})^{-1/2}=246$\,GeV.
The luminosity eventually reached values of up to $4 \cdot 10^{31}$\,cm$^{-2}$\,s$^{-1}$,
and a total  integrated luminosity  
of $0.5$\,fb$^{-1}$ was collected by H1 and as well by ZEUS in 15 years.
 HERA established the $ep$ scattering energy frontier and 
served to a large extent as a search laboratory for new physics
especially through configurations characteristic for the spacelike
$ep$ configuration such as leptoquarks or $R-$parity violating supersymmetry.
Albeit certain fluctuations appeared, especially on leptoquark and exotic multi-lepton 
signatures, one concluded that in the mass range 
explored by HERA new particles, with certain couplings, would not exist. 
H1 and ZEUS confirmed the unification of the electromagnetic and weak
interactions at 4-momentum transfers, $\sqrt{Q^2} \leq v$,
through high $Q^2$ measurements of  photon and $Z$ exchange in NC
and of $W$ exchange in CC~\cite{Abramowicz:2015mha}. HERA discovered
a rise of the sea quark, $x\Sigma$,  and gluon, $xg$, 
densities in the newly accessed region of small Bjorken $x$, by 
precisely measuring the structure function $F_2(x,Q^2)$ and its derivative
$\partial F_2/\partial \ln Q^2$, respectively. Despite some peculiarities
in the QCD analysis of the NC cross section at low $x$ and 
$Q^2$~\cite{Abramowicz:2015mha},
the linear $Q^2$ evolution DGLAP equations, the monumental
predictions of Guido Altarelli and Lev Lipatov with collaborators, were 
established to hold in the full HERA range which could be seen as indeed 
a surprise as the $\ln 1/x$ terms, characteristic to the BFKL evolution, were
large at small $x$. 
Often overlooked, but of
crucial relevance for future low $x$ physics, as $x \propto Q^2/s$,
we also discovered that the gluon distribution would not rise towards low $x$
but, on the contrary, 
disappear when $Q^2$ was approaching the mass of the proton squared region
$M_p^2 \sim 1$\,GeV$^2$, at the edge of the
DIS kinematic region. Following some initial EMC data on charm production in DIS,
HERA also measured the charm and bottom quark densities
in the proton and provided an experimental base for their theoretical study
in a variety of dynamic schemes of changing flavour numbers which now 
turn out to be quite relevant for the description of $pp$ data such as 
on associated $b$ and vector boson production at the LHC. 
Another striking observation at HERA was the finding that the proton remained
intact in about $10$\,\% of the violent $ep$ collisions which gave birth to
the field of diffractive DIS.  Instrumentation and physics at HERA were largely 
innovative, and many more important measurements were
done~\cite{Klein:2008di}. 

Seen from the LHC, 
HERA has been the indispensable, only base for adequately describing the parton-parton
interactions in $pp$, such as the gluon-gluon fusion to generate the Higgs boson.
The field of parton distribution functions, PDFs, was born 
which prior to HERA was at its infancy~\cite{Klein:2010zzc}.
For QCD, HERA demonstrated that the structure of matter is much richer than
the collinear approximation leading to the classic proton PDFs suggests.
Measurements of deeply virtual Compton scattering, diffraction, jets, photons, neutrons
opened new areas, supported by impressive theory developments,
for generalised, pomeron, unintegrated, photon,
pion and neutron structure functions. All of these ought to be studied much deeper
for understanding the universe at truly microscopic scales.


HERA kept electron-proton scattering as an integral part of high energy
particle physics. It demonstrated the 
richness of DIS physics and the feasibility of constructing and operating energy frontier $ep$ colliders.
It is a testimony of the vision and authority of Bjoern Wiik. What did we learn to take into
a next higher energy $ep$ collider design? Perhaps there 
were three lessons about i) the need for higher energy, for three reasons: 
to make  charged currents a real, precision part of $ep$ physics, for instance for the complete unfolding
of the flavour composition of the sea and valence quarks, to produce heavier mass particles (Higgs, top,
exotics) with favourable cross sections and, a third reason, to discover or disproof the existence of 
gluon saturation for which one needs to measure at lower $x \propto Q^2/s$ than HERA could; ii)
the need for much, much higher luminosity: the first almost ten years of HERA provided 
just a hundred pb$^{-1}$. As a consequence, HERA could not accurately
 access the high $x$ region, and it was inefficient and short of statistics in resolving 
 puzzling event fluctuations; iii) the complexity of the interaction region design when
 a bent electron beam caused synchrotron radiation while the opposite proton beam
 generated quite some halo background through beam-gas and beam-wall proton-ion interactions.
 This we had not seen clearly enough prior to and during the initial phase of the 
 HERA luminosity upgrade.

A key question  to the future of Deep Inelastic Scattering, to both its 
particle and nuclear physics components, is the choice of energy. This is an issue also in the 
debate~\cite{Aschenauer:2017jsk}
 about the current EIC designs which have a maximum of luminosity at a $\sqrt{s}$ of either
$35$\, 
(for the Jlab EIC~\cite{Abeyratne:2015pma}) 
or $100$\,GeV (for eRHIC~\cite{Aschenauer:2014cki}). 
This discussion, however, too often leaves the comparison with the LHeC 
 out\footnote{One here considers, for example, an extension of saturation
 scales $Q^2_s$ from $1$ to $2$\,GeV$^2$ while it is known that at so low
 $Q^2$ the gluon density at lower $x$ is practically zero, and at the edge of having
 a meaning anyhow since one is near the limit of application of pQCD.
 A true discovery of the saturation of a rise of $xg$
 requires to be at $very$ low $x$ where densities are large but also
 in the perturbative region
 where $\alpha_s$ is small, i.e. $Q^2 \geq 10$\,GeV$^2$. One cannot discover
 a saturating gluon density in a region where it is non-existing and no
 nuclear environment amplification, as speculative as it is by itself, can
 change that conclusion.}, which has a $\sqrt{s}$ of $1300$\,GeV.
   A complete view is desirable, extending first attempts such as~\cite{mkpoetic16}, as to what 
 can be learned from going back by one or two orders of magnitude in $Q^2$ below HERA, as
 with  the US EICs, 
 and what can be achieved by a large increase, as with LHeC or even FCC-eh, 
 and, furthermore, which energy 
 range was crucial to be covered in electron-ion scattering. It would then become clearer
 as to where real synergy resides. A particular question regards the puzzle on the composition of the
proton spin which at HERA was investigated by the HERMES experiment, and has recently
been claimed to be resolved in lattice QCD~\cite{Alexandrou:2017oeh}. That does $not$
require maximum energies, as the asymmetries of polarised scattering cross sections 
to measure vanish proportional to $x$. Very high energy in $eh$ scattering is 
crucial for a DIS Higgs, top, BSM, PDF, vector meson  and low $x$ programme.
One crudely may call the low energy EICs extended continuations of the 
NMC, COMPASS and HERMES research programs, on nuclear structure, 
QCD and spin, while the LHeC is in the line of H1 and ZEUS with the important
extension to electron-deuteron and electron-ion scattering which HERA was not
able to study.    A question is whether the global nuclear and particle 
physics communities have the intention and influence to shape a common future 
of DIS which HERA has opened and which can only be rich and convincing if the programme as
a whole is considered.

\section{Lessons from the LHC}
\subsection{Proton-Proton Scattering}
Expectations for the observation at the LHC of new physics beyond the Standard Model
had been almost unlimited as is illustrated, for example,  by the comprehensive 
study of the ATLAS Collaboration on its physics potential which appeared prior 
to data taking~\cite{Aad:2009wy}. And yet, the first sizeable data sets revealed no
new particle nor  symmetry.
 At the EPS Conference 2011 at Grenoble, Guido Altarelli, in a memorable 
plenary talk, stated it was ``too early for desparation but enough for despair". In 2012
ATLAS and CMS discovered the Higgs boson, termed the last and a most crucial 
component of the SM gauge field theory. As more data had been collected and
analysed (at the time of this article LHC delivered a record integrated luminosity
of $100$\,fb$^{-1}$) Guido later noted: {\it{It is now less 
unconceivable that no new physics will show up at the 
LHC... We expected complexity and instead we have found a maximum of simplicity..
The possibility that the Standard model holds well beyond the electroweak 
scale must now be seriously considered~\cite{Altarelli:2014xxa}.}} 

If today one talks about lessons 
from the LHC one may emphasise the following observations: i)~the LHC and the 
general purpose detectors ATLAS and CMS, which just celebrate the 25 years since the
approval of their letters of intent, and also the other experiments as LHCb, have operated
with high efficiency and reliable success owing to the ingenuitive work of thousands of 
physicists and engineers. The LHC and its experiments comprise the overwhelming
majority of all HEP efforts and had they not succeeded, particle physics would
be in danger of its existence; ii) there has been no observation of supersymmetric 
particles of mass up to O(1)\,TeV and no observation of any of the considered
extensions of the SM, such as  Kaluza Klein excitations of gravitons or 
new vector bosons $Z'$ or $W'$, currently up to masses of O(5)\,TeV. The absence
of physics BSM~\footnote{This has had
major consequences especially for the layout of new electron-positron colliders:
at CERN and in China circular machine designs reappeared when it was
most likely that an electron-positron collider above the scale set by Higgs and top production
was an ``empty machine". Interesting enough, both the Chinese CepC and the
most recent version of the Japanese ILC set $\sqrt{s}=2E_e \simeq  v$, in order
to measure the Higgs-strahlung  from $Z$, $e^+e^- \rightarrow Z^* \rightarrow ZH$.
The coming energy frontier in $e^+e^-$ is nothing but the Fermi scale
with a question mark about CLIC which could principally reach $3$\,TeV
at high cost should a new spectrum of particles still be discovered at subsequent
LHC runs or/and new theories inevitably led to the conclusion that new physics
was in reach at a few TeV mass.}   had been the biggest surprise at the LHC so far; iii)
a next lesson at the LHC was the confirmation of the SM at a striking level of
precision, if one considers the measurement of the inclusive Drell-Yan scattering
cross sections of the $W$ and $Z$ bosons \cite{Aaboud:2016btc} to about half a percent
precision, apart from the also striking $2$\,\% luminosity uncertainty,  or the
recent measurement of the mass of the $W$ boson to $19$\,MeV 
precision~\cite{Aaboud:2017svj}; iv) QCD and electroweak theory had been extended
to high orders, with the N$^3$LO calculation of the Higgs production at the LHC
as an outstanding example. The success of the LHC is its challenge. Globally
a huge effort is undertaken to upgrade the LHC accelerator and its experiments
to highest reasonable, in terms of event pile-up, luminosities. The HL LHC
is projected to operate for two more decades hence, and a fundamental question is that about its science base. The addition of $ep$ adds new possibilities for the discovery
of physics beyond the SM $and$ it transforms the LHC facility eventually into
a precision particle physics laboratory of wider and deeper range than
a sole $e^+e^-$ machine can possibly represent.
\subsection{Scattering of Heavy Ions}
The $PbPb$ runs at the LHC have confirmed the picture coming from RHIC that the
medium created in such collisions behaves very early like a very low viscosity
fluid whose behaviour can be well described by relativistic viscous
hydrodynamics. It is very opaque to particles traversing it, a phenomenon known
as jet quenching. The first statement is supported by extensive studies of
azimuthal asymmetries far beyond, both in kinematic reach and order of
correlations, those that could be studied at RHIC. The second one is provided by
a wealth of measurements, including jet yields and shapes both isolated and
accompanied by an electroweak boson, which 
have triggered a huge theoretical activity to understand the modifications
of the QCD branching inside a coloured medium, 
for a review see~\cite{Armesto:2015ioy}. Finally, a great surprise came
from the measurements of smaller systems, $pp$ and $pPb$,
which revealed a not expected smooth transition
when increasing the colliding system size for observables (ridge and
azimuthal asymmetries, particle species, interferometry,...) that have been
traditionally considered as signatures of the creation of deconfined partonic
matter close to thermodynamical equilibriumÊ - Quark Gluon Plasma - in
high-energy nucleus-nucleus collisions. In short, all hadron-hadron 
collisions at the LHC were observed to be of collective nature. It  therefore
needs the electron-ion configuration to i) establish a new level of 
understanding the parton dynamics inside nuclei in the kinematic
range covered by the HI experiments at the LHC
  and to ii) provide the theoretically sound initial conditions
for the macroscopic evolution of the system leading eventually to
an understanding of the emergence  of that hydrodynamical 
macroscopic behaviour from the QCD microscopic dynamics.

\section{The LHeC Project}
Owing to the high energy, intense hadron beams of the LHC, CERN has the 
unique potential for discovery and precision DIS at the energy frontier and 
{\it{it would be a waste}} not to utilise that as Guido had 
formulated already ten years ago~\cite{guido08}. The idea of $ep$ collisions
with the LHC was considered already at the first LHC workshop in 1984,
again at the Aachen workshop in 1990, and $ep$ was presented {\it{as a bonus}}
when Chris Llewellyn Smith, then Director General of CERN, 
announced the approval of the LHC now 25 years ago,
for a brief account of $ep$ collider history see the introduction to~\cite{Abelleira2012}.
The idea was revived  when CERN, ECFA and later NuPECC invited a study of 
how one may realise $ep$ and $eA$ collisions in view of the LHC physics findings 
and accelerator and detector technology advances. This was the base
for the extended conceptual design report (CDR) of the LHeC~\cite{Abelleira2012}. 
It was published in summer 2012. At just that time the Higgs boson was discovered.
The CDR had presented a machine of $10^{33}$\,cm$^{-2}$s$^{-1}$ luminosity,
about a hundred times more intense than HERA I. 
The CDR had recognised~\cite{Abelleira2012} that the $H$ boson, 
predominantly produced in CC DIS, was generated 
with a cross section of $200$\,fb in polarised  $ep \rightarrow \nu H X$
as large as in $e^+e^-$. The $H$ production is through radiation from
a spacelike $W$ (CC) or $Z$ (NC), two production mechanisms
 which unlike in $pp$ are uniquely distinguishable.
The final state is clean as there is no pileup at the LHeC. This caused
a reconsideration of the accelerator design assumptions, especially a modest
enlargement of the electron beam current and reduction of the $\beta^*$ parameter. 
Since, moreover, the beam brightness, i.e. the ratio of number of protons per bunch
and the proton beam emittance, was found to be almost three times larger
than the so-called ``ultimate beam parameters",  predicted at that time to guard the 
future LHC and adopted in the LHeC CDR, a path
had been found for the LHeC to become a genuine and competitive Higgs
physics facility at atobarn reach of integrated 
luminosity~\cite{Zimmermann:2013aga,Bruening:2013bga}. 
It is interesting to recall that Guido Altarelli, in all his LHeC presentations,
nevertheless emphasised the high precision, large kinematic range
character of the genuine DIS programme, see below, which he knew to value so much 
that Higgs and BSM, as interesting as they could be, were not required by him to justify
an $ep/eA$ energy frontier experiment at CERN~\footnote{This often reminded
me on the proposal for what later became the quark discovery $ep$ experiment
at SLAC~\cite{Bloom:1969kc,Breidenbach:1969kd}
which had stated the main aim would be ``to collect data which could
be of use for future experiments". One is struck by such reasons when nowadays
no less than a discovery of something like Dark Matter is requested for a project
to proceed. The two mile SLAC linac was longer than the two times 1\,km linacs
of the LHeC, but the physics program requirements are orders of magnitude
harder to satisfy. One is tempted to note certain arrogance of ourselves towards
nature and its predictability. The present ``no BSM physics" situation of the
LHC may lead to a reset of our ambitions and expectations towards a more
humble understanding of what we may expect and what we should promise.}. 
%
\subsection{Principal Design Considerations}
The LHeC should provide an intense, high energy electron beam 
to collide with the LHC. The intensity is gauged through the luminosity 
goal of O($1$)\,ab$^{-1}$, the energy chosen to achieve a TeV energy
collision. At the time of this write up, a cost-physics-energy evaluation
takes place with the goal to be able to realise the LHeC with a 
most far reaching physics scope. 
 The wall plug power of the default design~\cite{Abelleira2012}  had been
constrained to $100$\,MW. In that configuration
two super-conducting linacs, opposite to each other,
accelerate the passing electrons by $10$\,GeV each. This
leads to a final electron beam of $60$\,GeV in a 3-turn racetrack configuration
as   is illustrated in Fig.\,\ref{figlhec}.  For measuring at very low $Q^2$ 
and/or determining $F_L$, see below, the electron beam energy may be reduced to 
a minimum of perhaps $20$\,GeV. For maximising the acceptance at large Bjorken
$x$, the proton beam energy, $E_p$, may be reduced to $1$\,TeV about. If the ERL
may be combined in the further future with a double energy (HE) LHC, $E_p$
could reach $14$\,TeV. One therefore considers a DIS scattering complex with an
energy range between $\sqrt{s} \simeq 300$ to $1800$\,GeV. It thus
covers a range from HERA to the TeV region, at hugely increased luminosity
and much more sophisticated experimental techniques.

The ERL arrangement, Fig.\,\ref{figlhec}, is located inside the LHC ring but
outside its tunnel, which minimises any interference
with the main hadron beam infrastructure. 
The electron accelerator may thus
be built independently, to a considerable extent, of the status of operation
of the proton machine.  This is the principal advantage of the Linac-Ring
over the Ring-Ring configuration which had been studied to considerable 
detail in the CDR also.

The chosen energy  of $60$\,GeV
leads to a circumference $U$ of the electron
racetrack of $8.9$\,km. This length is a fraction
$1/n$ of the LHC circumference, for $n=3$, as is required 
for the $e$ and $p$ matching of bunch patterns. 
It is chosen also in order to limit the energy loss in the last return arc and as a result of
a cost optimisation between the fractions of the circumference covered by SRF 
and by return arcs. That  configuration is adopted also as the default  for the FCC-he. Tentatively,
for the LHC the ERL would be tangential to IP2 which, according to the current 
plans, is taken by the ALICE detector until the first long shutdown following
the three year pause of the LHC operation for upgrading the luminosity performance
and detectors. This shutdown is termed LS4 and may begin in 2030. For FCC-he
the preferred position is IP L for geological reasons mainly.

The LHeC operation is transparent to the LHC collider experiments owing to
the low lepton bunch charge and resulting small beam-beam tune shift
experienced by the protons. The LHeC is thus designed to run simultaneously with
$pp$ (or $AA$) collisions. This tames  the cost and optimises the
physics return since a concurrent operation with the LHC will have a direct
impact on the HL LHC physics programme as is sketched subsequently.

%
\begin{figure}[htbp]
\vspace{0.3cm}
\centerline{\includegraphics*[width=120mm]{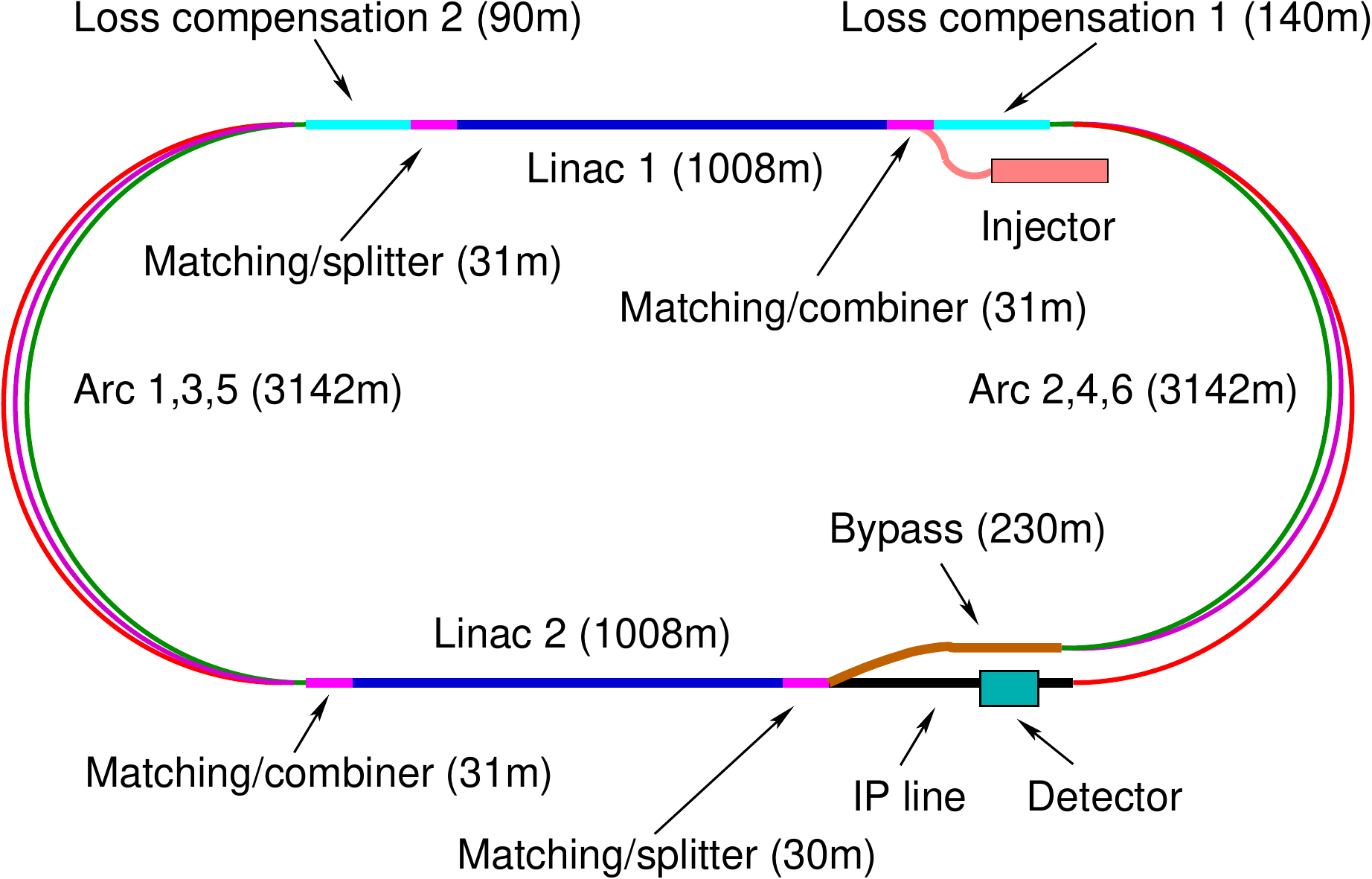}}
\vspace*{0.2cm}
\caption{\footnotesize{
Schematic view of the default LHeC configuration.
Each linac accelerates the beam to $10$\,GeV, which leads to a
$60$\,GeV electron energy at the interaction point after three
passes through the opposite lying linac structures made
of $60$ cavity-cryo modules
each. The cryomodules comprise four 5-cell cavities at $802$\,MHz frequency.
The arc radius is about $1$\,km,
and the circumference chosen to be $1/3$ of that of the LHC.
The beam is  decelerated for recovering the beam power
after having passed the IP. The dump therefore is very ambient friendly.
}}
\label{figlhec}
\end{figure}
\subsection{Parameters}
Following~\cite{fccnote} the parameters of the current LHeC default design 
may be summarised as follows.
The luminosity $L$ of the linac-ring $ep$ collider may be approximated as 
\begin{equation}
L  = \frac{N_p N_e f \gamma_p}{4 \pi \epsilon_p\beta_p} \cdot H_{geo}H_{bb}H_{fill}.
\end{equation}
Here, $N_p$ is the number of protons per bunch and 
$\epsilon_p$ and $\beta_p$ are the proton emittance and beta-functions. 
The main proton beam parameters $N_p$ and $\epsilon_p$ are defined
by the LHC GPD experiments  in concurrent $ep$ and $pp$ operation. 
The proton beta-function in the electron-proton collision point  may be as small
as $\beta_p=7\;\rm cm$.
\begin{table*}[h]
\begin{center}
{\begin{tabular}{|l|c|c|c|c|}
\hline
parameter [unit] & LHeC CDR & ep at HL-LHC & ep at HE-LHC & FCC-he \\
\hline
$E_p $ [TeV] &  7 &  7 &  12.5 &  50 \\ 
$E_e$ [GeV] & 60 & 60 & 60 & 60 \\
$\sqrt{s}$ [TeV] & 1.3 & 1.3 & 1.7 & 3.5 \\
bunch spacing [ns] & 25 & 25 & 25 & 25 \\ 
protons per bunch [$10^{11}$] & 1.7 & 2.2 & 2.5 & 1 \\
$\gamma \epsilon_p$ [$\mu$m] & 3.7 & 2 & 2.5 & 2.2 \\
electrons per bunch [$10^9$] & 1 & 2.3 & 3.0 & 3.0 \\
electron current [mA] & 6.4 & 15 & 20 & 20 \\
IP beta function~~$\beta^*_p$ [cm] & 10 & 7 & 10 & 15 \\
hourglass factor~~H$_{geo}$ & 0.9 & 0.9 & 0.9& 0.9 \\
pinch factor~~H$_{bb}$ & 1.3 & 1.3 & 1.3 & 1.3 \\
proton filling~~H$_{fill}$ & 0.8 & 0.8 & 0.8 & 0.8 \\
luminosity [$10^{33}$cm$^{-2}$s$^{-1}$] & 1 & 8 & 12 & 15 \\
\hline
\end{tabular} \label{tabpar}}
\caption{\footnotesize{Baseline parameters and estimated peak luminosities
of future electron-proton collider configurations 
for the electron ERL when used in concurrent $ep$ and $pp$ operation mode,
taken from~\cite{fccnote}.}}

\end{center}
\end{table*}
Furthermore, $f = 1/\Delta$  denotes the bunch frequency, which for
the default  bunch spacing of $\Delta=25$\,ns  is= $40$\,MHz.
$N_e$ is the number of electrons per bunch which determines the electron current
$I_e =e N_e f$ with a target value of $15$\,mA, a twofold increase compared to
the conservative CDR.  For $60$\,GeV, this will yield a total synchrotron radiation of
about $40\;\rm MW$ in the return arcs. To compensate for this power 
loss through the beam, a  grid power of the order of $65\;\rm MW$ may be required. This load
may be substantially reduced if the electron energy is diminished.

The factors $H_{geo}$, $H_{bb}$ and $H_{fill}$ are geometric correction
factors with values typically close to unity.
$H_{geo}$ is the reduction of the luminosity due to the hourglass effect, 
$H_{bb}$ is the increase of the luminosity by the strong attractive beam-beam forces and
$H_{fill}$ is a factor that takes the filling patters of the electron and the proton beam into account. 
Estimates for these parameters are shown in Tab.\,\ref{tabpar}. 
Compared to the CDR of the LHeC from 2012, it seems indeed
possible to achieve peak luminosities near to or larger than 
$10^{34}$\,cm$^{-2}$s$^{-1}$, which makes these future $ep$ colliders 
most exciting and efficient machines for the study of new physics 
at the accelerator energy frontier.
\subsection{The Energy Recovery Development Facility PERLE at Orsay}
When Guido Altarelli had asked, at the 2015 LHeC workshop, whether we could 
build the LHeC, the IAC, of which he was a very prominent member, discussed 
the importance of developing the basic technology of the LHeC. Since then
a conceptual design report was published~\cite{Angal-Kalinin:2017iup}
of a low energy, energy recovery linac (ERL) facility. 
A collaboration has been founded with 
the intent to realise a ``Powerful Energy Recovery Linac for Experiments" (PERLE)
at the site of LAL Orsay where a hall and parts of the infrastructure
already exist. Initially participating  
institutes are BINP Novosibirsk, CERN, Daresbury, the University of Liverpool,
Jefferson Laboratory and the two now merging Orsay Institutes, LAL and IPN.
The main parameters of PERLE are derived from the LHeC, such as the electron
current of $20$\,mA chosen to achieve its $10^{34}$\,cm$^{-2}$s$^{-1}$ luminosity
target.

The PERLE layout considers two  accelerating superconducting cavity structures, 
cryogenic modules each housing four $802$\,MHz frequency $5$-cell cavities, which are
embedded at opposite linear parts in
a racetrack configuration with two triple arcs. Operation for ERL
is in CW mode at typical gradients of $18$\,MV/m. An electron beam of up to $20$\,mA current,
derived from a DC photocathode, is injected with $7$\,MeV energy into the PERLE
lattice which is an ellipse of main dimensions $5.5 \times 24$\,m$^2$. With three turns
the beam is accelerated to about $500$\,MeV and, after a $\pi/2$ phase shift, decelerated 
and dumped at the injection energy. PERLE therefore represents a unique facility
with its multi-turn ERL and a new, high power level of $10$\,MW which distinguishes
it from various other facilities at lower energy or different current, frequency
or return arc magnet technology.

\begin{figure}[htbp]
\vspace{0.3cm}
\centerline{\includegraphics*[width=120mm]{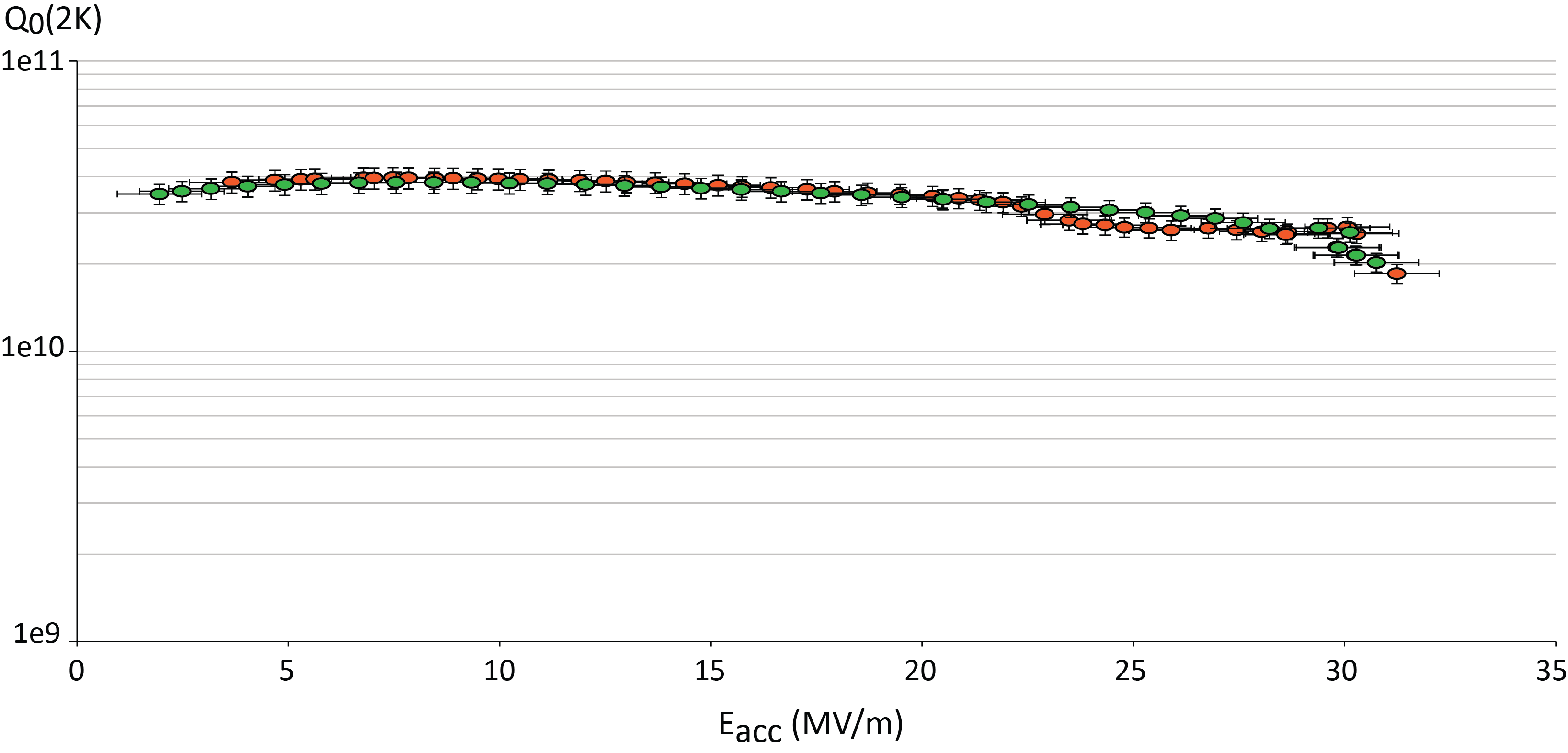}}
\vspace*{0.2cm}
\caption{\footnotesize{
Quality factor $Q_0$ of the first $5$-cell, $801.58$\,MHz frequency, Niobium, PERLE prototype
cavity determined as a function of the acceleration gradient. The observed behaviour
surpasses by far the expectation, with a $Q_0$ well exceeding $10^{10}$ and an apparent
very weak degradation towards gradients extending much beyond the considered CW operation 
working point of $18$\,MV/m, from~\cite{frankcavity}.
}}
\label{figq0}
\end{figure}

The frequency of $802$\,MHz is chosen to comply with the LHC constraint following 
an optimisation involving beam breakup stability for the LHeC, cost of the structure
and RF power as well as cryogenic capacity~\cite{erk,frank}. A $5$-cell Niobium
cavity prototype for PERLE, designed by CERN and Jlab~\cite{Angal-Kalinin:2017iup}, 
has been built at Jefferson Laboratory in 2017 and successfully passed all checks for 
mechanical stability, tolerance and flatness. Very recently its $Q_0$ factor, 
which determines the maximum of stored
energy relatively to the input power, has been initially measured and was observed to
safely surpass the requirement of a value in excess of $10^{10}$, see Fig.\,\ref{figq0}. 
The development of PERLE is thus passing
its first and a crucial milestone, with further investigations
as on the field emission to possibly follow and elements for higher order
mode extractions under design.

Currently the PERLE Collaboration works on the technical design of the facility with
emphasis on building the first cryomodule, using the considerable experience gained with 
the existing SPL, ESS and SNS modules developed by CERN, IPN Orsay and Jlab, respectively.
 It is hoped that PERLE can start operation within a few years time, also building on the large DC 
 photocathode experience at Daresbury Lab and on the art of magnet fabrication at 
 BINP Novosibirsk. Primarily, PERLE is built as an accelerator R+D facility. Questions such 
 as on the time structure, transverse and longitudinal matching, source limitations, space charge,
beam break up, halo and other effects ought to be investigated in order
to master ERL technology and dynamics at the $10$\,MW power level. 

The high intensity and chosen energy make PERLE in the mid term future
a unique facility for the study of photo-nuclear reactions
and also particle and nuclear electron beam physics on, for example, the
study of the scale dependence of the weak mixing angle, a precise
measurement of the proton radius or the search for dark light 
photons. Other applications, also described in the PERLE CDR~\cite{Angal-Kalinin:2017iup},
comprise the beam based development of SCRF, the test of accelerator components, 
such as the quench behaviour of superconducting magnets, or detector developments. 
With a photon intensity exceeding by orders of magnitude that of the
ELI facility~\cite{eli}  currently under construction in Southern Europe,
PERLE represents a far reaching new accelerator base for the laboratories
in the Orsay and Saclay area, the further
collaborating institutes and future possible users for fundamental and applied
physics investigations. PERLE may also be configured as a powerful laser physics
facility when complemented with an undulator system. At the same time, experience and technology emerging from
PERLE are enablers for the LHeC, the highest energy ERL application so far considered.
\subsection{Detector}
The arrangement of the electron ERL tangential to the LHC defines one interaction
region (IR) for one detector. By LS4, now scheduled for 2030, the ion-ion program
of the LHC ends in order to maximise the $pp$ luminosity collection,
and currently the operation of the ALICE detector at IP2 is considered to then be complete.
This opens the opportunity for replacing it with a new detector for precision
electron-hadron physics. A study has been made~\cite{gaddi} of the sequence of 
installing the new detector in IP2. Based on experience with the pre-mounting 
of the CMS detector and exploiting the modular structure of the LHeC detector 
it was established that a time of less than 2 years shall be sufficient for such
an ``exchange of detectors". Typical shutdown periods of the LHC range from 
one to nearly three years, and one may expect that the first long shutdown after the upgrade of
machine and detectors may no be short for reasons independent of $ep$. This
opens the real opportunity for DIS physics with the LHC in the thirties.

\begin{figure}[hb]
\vspace{-.4cm}
\centerline{\includegraphics*[width=140mm]{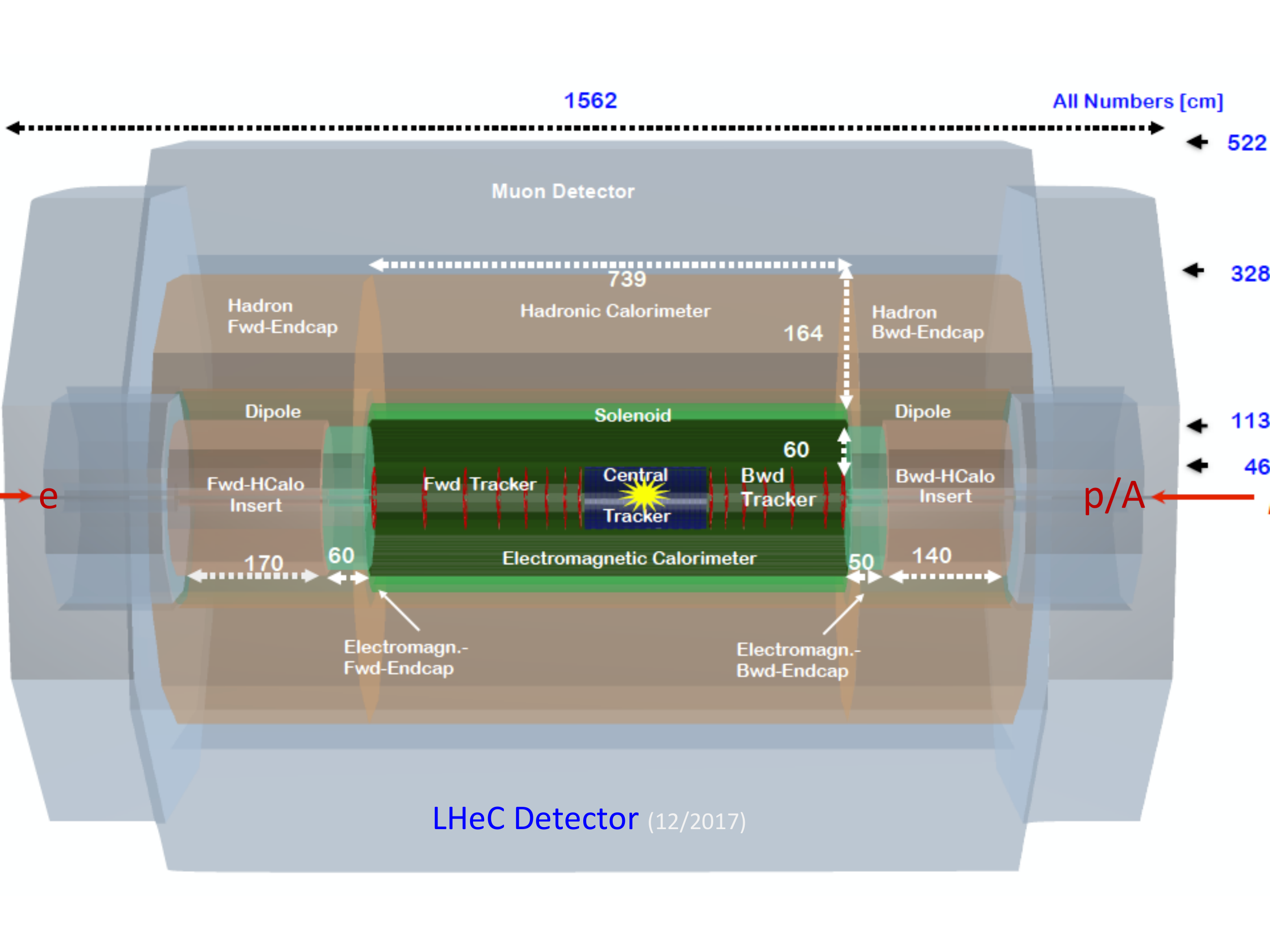}}
\vspace*{-0.1cm}
\caption{\footnotesize{
View on the LHeC detector under design to operate with both the high luminosity (HL)
and the high energy (HE) LHC. The proton beam comes from the right, pointing to
 the forward region to which jets and electrons may be scattered of multi-TeV
 energies. The electron beam points into the backward region which has to measure
 hadron and electromagnetic energies not exceeding the electron beam energy. The detector
 is therefore asymmetric. Special attention is required for forward calorimetry
 and the polar angle acceptance, essential for forward $b$ tagging and for low $Q^2$
 backward scattered electrons. The beam pipe is eccentric to permit the synchrotron
 radiation fan of the bent electron beam to pass through the interaction region.
  The HE (HL) central LHeC detector shown here has a length of 
 $15.6~(13.2)$\,m and a radius of $5.2~(4.4)$\,m. This may be compared with the 
 CMS dimensions of $21 \times 7.5$\,m$^2$.   
}}
\label{fighedetnr}
\end{figure}

In considering the further future one may expect that the LHC be replaced by 
its double energy version, the HE LHC, following
the termination of the LHC operation currently envisaged to take place in $\sim 2038$ 
after accumulating $3$ or $4$\,ab$^{-1}$ of integrated $pp$ luminosity.
It thus is conceivable that the $ep$ detector may also be used for operation during the
HE LHC phase, from likely the end of the forties to two decades hence. It then
seemed not illogical to design an $ep$ detector which could cope with first $7$ and later
$\simeq 13$\,TeV proton beams. 
The current status of its design is sketched in Fig.\,\ref{fighedetnr}.
The detector has a classic collider detector structure and its basic design 
considerations as well as simulations are summarised in the LHeC 
CDR~\cite{Abelleira2012}. Its characteristic feature is a combined magnet structure 
which has to be inserted between the electromagnetic and hadronic calorimeters,
i.e. a dipole magnet for bending the electron beam into and out of the proton beam axis
and a $3$\,T solenoid currently combined with that dipole. Two versions exist of warm and 
LAr electromagnetic calorimetry.

Recent work  focusses on the  adaption to 
evolving detector technology, such as low material, high resolution Silicon detectors,
mainly in connection with the LHC luminosity upgrade. The LHeC conditions are 
comparatively easier  with an $ep$ event pile-up of only $0.1$
and an about hundred times lower radiation level compared to the LHC. Strong efforts 
are made to update the IR design from the CDR to new types of quadrupoles
of which high field $Nb_3Sn$ actively shielded quadrupoles are currently
most promising~\cite{brett}. The design of the IR is most challenging
and at the end of 2017 still work in progress in collaboration between
BNL, CERN and Liverpool. Much care and time are also dedicated to  
the development of DD4HEP/DDG4 based design, simulation and reconstruction 
software.  The central LHeC detector must be completed
by forward proton, neutron and possibly deuteron taggers while the backward
region has to be instrumented with near axis photon and electron detectors to
measure the $ep$ luminosity from Bethe-Heitler scattering and to tag and normalise the
photoproduction background to genuine DIS. These taggers have all been considered 
in the CDR.
It is intended to update the CDR detector design within 2018 and to
possibly assemble an LHeC Collaboration which would certainly move these 
design concept studies, currently also involving service routing and material 
minimization, much forward.
\section{Resolving the Substructure of Matter}
DIS is the means to explore the substructure of matter using photons and $W,~Z$ 
bosons as probes with the kinematics
fixed by the scattered electron (NC) or neutrino (CC). Both the (negative) $4$-momentum 
squared, $Q^2$, and the parton momentum fraction $x$ are
determined externally to the virtual $\gamma$ ($W$ or $Z$) interaction with the  parton. 
The kinematics can equally well be determined with the hadronic final state. The resolution 
is $1/\sqrt{Q^2}$, where $1$\,GeV corresponds to a distance of $0.2$\,fm, which is how 
Hofstatter discovered a finite proton radius of about $0.8$\,fm with an electron beam of $200$\,MeV. 
Following a series of fixed target lepton scattering DIS experiments, with $Q^2 \leq s=2M_pE_l$, 
HERA made a big step forward by increasing the cms energy squared $s=4 E_eE_p$ to 
$10^5$\,GeV$^2$ in the first $ep$ collider ever built. That range could not be fully explored 
because of the low luminosity of HERA. The LHeC is designed to achieve a factor of $1000$  
higher integrated luminosity, and extends the energy squared to $1.7 \cdot 10^6$\,GeV$^2$. 
This will lead to quantitatively more valuable and much deeper reaching insight into the 
structure and dynamics of parton interactions inside matter.
\subsection{The Proton's Partonic Structure and Dynamics}
\label{secpdf}
The momentum distributions of partons inside the proton, $xP(x,Q^2)$, must be determined from experiment. The most suited process to obtain an unbiased determination of the quark and gluon distributions  is inclusive deep inelastic NC and CC scattering. ``PDFs" as they are now termed have two meanings in that i) they represent a probability view on the substructure of the proton, or similarly of other hadrons, at a given distance $1/\sqrt{Q^2}$, and ii) they   describe, supposedly universally 
through cross section factorisation theorems~\cite{Collins:1989gx}, the hard scattering processes involving partons~\footnote{In his referee report on the LHeC CDR, in 2012, Guido Altarelli noted on the factorisation theorem in QCD for hadron colliders that: {\it{many
people still advance doubts. Actually this question could be studied
experimentally, in that the LHeC, with its improved precision, could put
bounds on the allowed amount of possible factorisation violations (eg by
measuring in DIS the gluon at large x and then comparing with jet
production at large pT in hadron colliders).}}}. Their determination therefore is crucial both for QCD and for searches for new states which makes them so fashionable nowadays at the LHC. DIS is $the$ process to determine PDFs, unlike Drell-Yan scattering, because it is theoretically clean, free of colour interactions in the initial or final state, and it has the prescribable variation of $Q^2$ and finally the determination of $x$ external to the intrinsic scattering process. 

The PDF programme of the  LHeC, as is briefly illustrated below, is of unprecedented depth for the following reasons: 
\begin{itemize}
\item it will resolve the partonic structure of the proton (and nuclei, see Sect.\,\ref{seceA}) for the first time completely, i.e. determine $P=u_v,~d_v,~u,~d,~s,~c,~b$ and the top and gluon momentum distributions through NC and CC cross section and direct heavy quark measurements in a huge kinematic range, from below $x=10^{-6}$ to $x=0.9$ and in $Q^2$ from below the DIS region to almost $Q^2=4E_eE_p=1.7 \cdot 10^6$\,GeV$^2$; 
\item a thousand-fold increase of the HERA luminosity, unprecedented precision from new detector technology and the redundant evaluation of the event kinematics from the lepton and hadron final state components will lead to extremely high precision and, technically important, to the fixation of the various PDF analysis parameters from the LHeC data themselves; 
\item because the high energy CC, unlike at HERA or low energy EICs, becomes equivalent to NC and thus no other data will be required: that is, there is no influence from higher twists or nuclear uncertainties, and no other experiment needed, i.e. LHeC will be the unique base for PDFs, for predictions, discovery and novel tests of theory.
\end{itemize}
Given the impressive theoretical progress on pQCD, see e.g.~\cite{Moch:2017uml}, one will have these PDFs available at N$^3$LO
and only then have the PDFs available to respond to the N$^3$LO Higgs in $pp$ cross-section calculations demand. For QCD, this will resolve open issues (and probably creating new ones) on $\alpha_s$, discussed in the Appendix, answer the question on the persistence (or not) of the linear parton evolution equations at small $x$, see Sect.\,\ref{seclowx}, and also decisively test whether factorisation holds or not between DIS and Drell-Yan scattering.

\begin{figure}[h]
\vspace*{-.1cm}
\centerline{\includegraphics*[width=75mm]{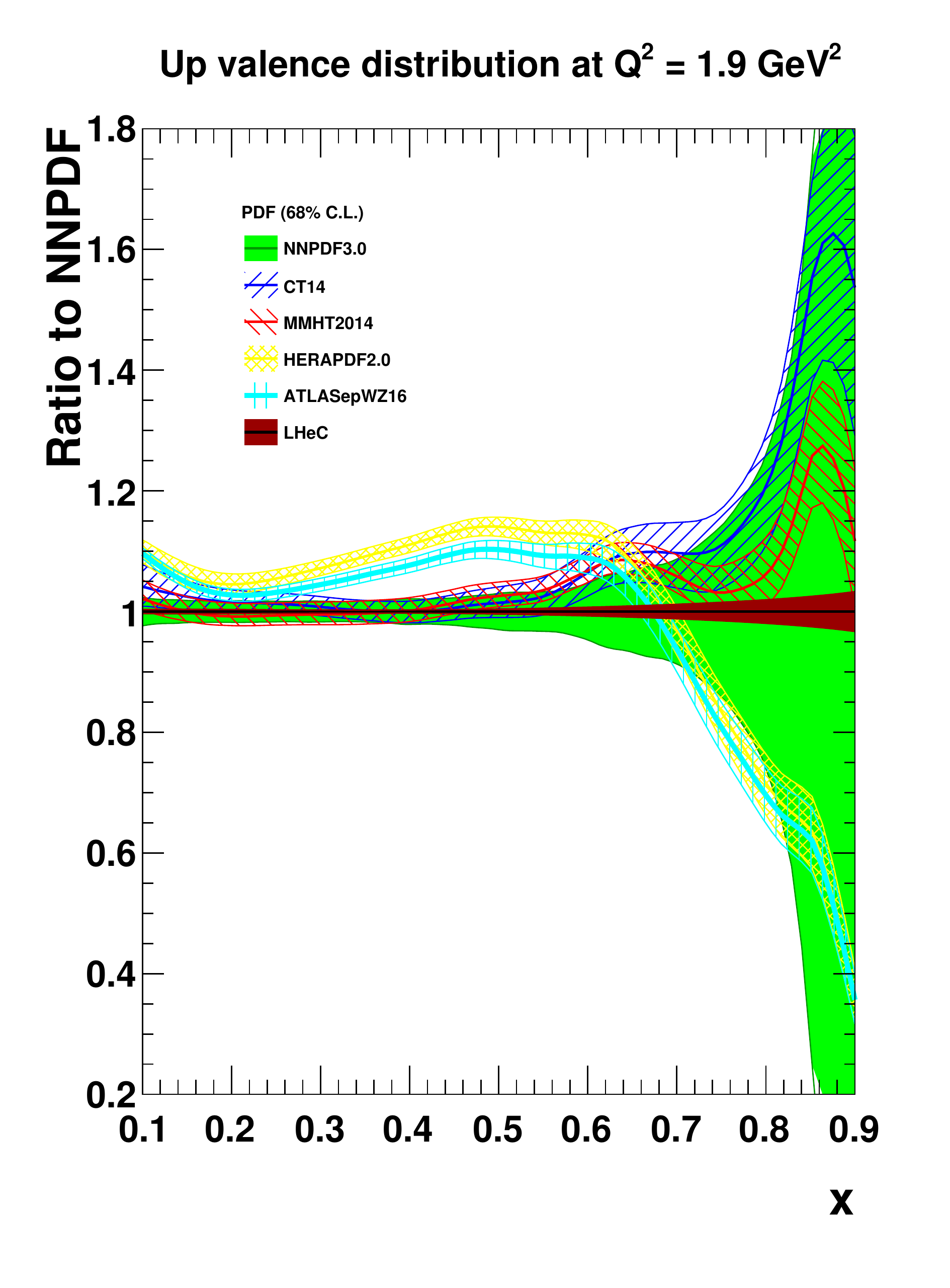}
\includegraphics*[width=75mm]{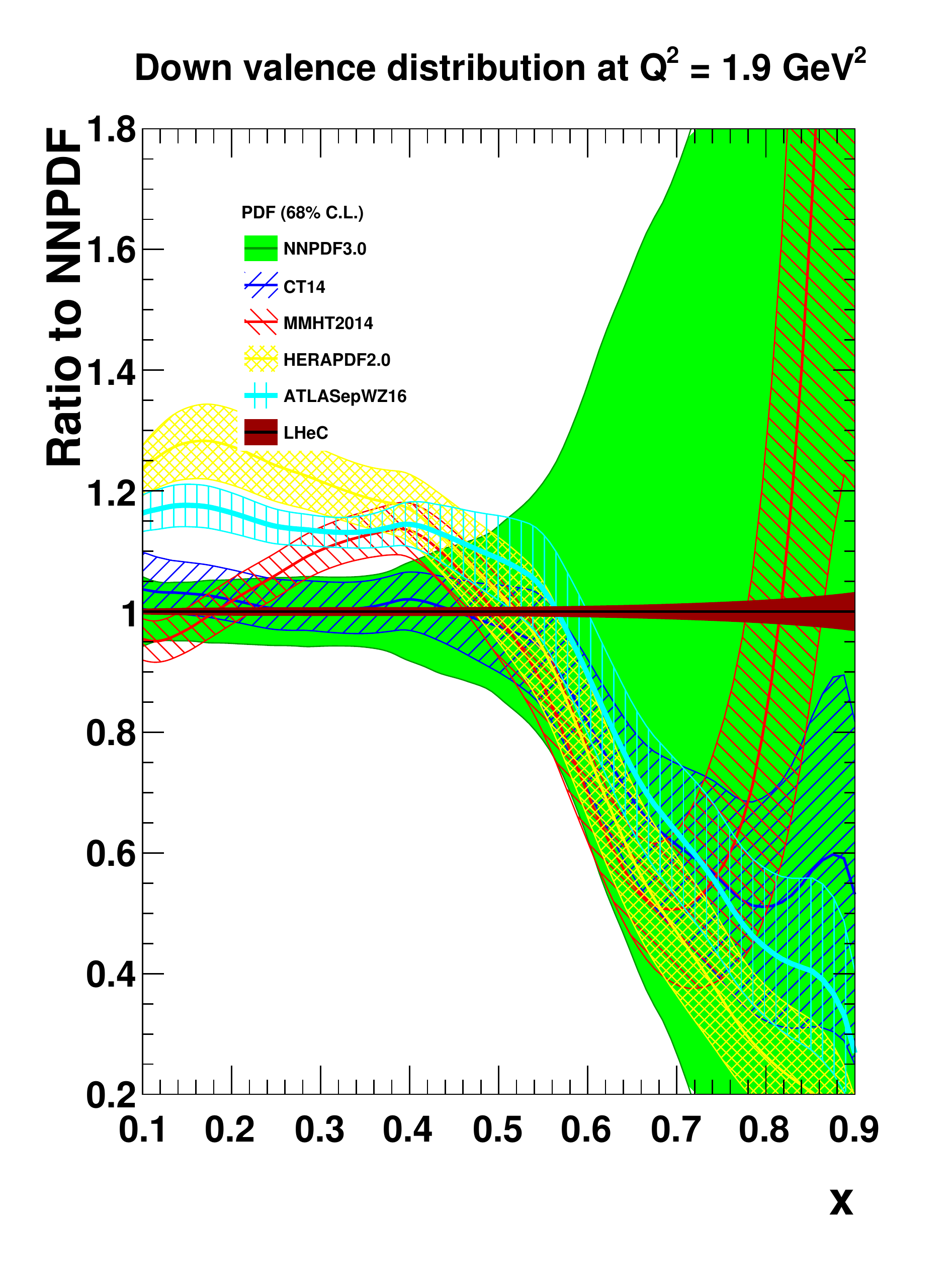}}
\vspace*{-.9cm}
\caption{\footnotesize{Determination of the valence quark distributions as functions of Bjorken $x$. Plotted are the ratios to the NNPDF result with uncertainties displayed as are provided by the individual sets,  left for the up-valence quark and right the down-valence quark distribution. For the LHeC the total uncertainty is plotted and the central value assumed to agree with NNPDF. As non-singlet quantities, the valence quark distributions are approximately the same with varying $Q^2$.}}
\label{figudv}
\end{figure}

The prospect for the determination of the proton-parton distributions with the LHeC has been studied in detail. 
It relies on the careful simulation of NC and CC measurements, including a full account of their systematic uncertainties and  expected correlations, as described in~\cite{Abelleira2012,voimax13}. Two examples are given on the expected precision. Surprising after 50 years of DIS, there is no certainty about the behaviour of the valence-quark distributions at large (and small) $x$. This is illustrated in 
Fig.\,\ref{figudv} which displays the uncertainties and central values for the up and the down valence quarks as a function of $x$ as are obtained in recent PDF  
analyses. One finds that the up-valence quark is better determined than the down-valence quark, which is related to the proton structure function $F_2/x$ having the large $x$ limit of $4u_v+d_v$. However, at large $x \geq 0.5$ where the distributions decrease, nuclear corrections rise and higher twist effects enter most, we are far from understanding the valence quark distributions. Moreover, there is an apparent difference of $\simeq 20$\,\% between various expectations even at medium $x$. The LHeC will determine both $u_v$ and $d_v$ with very high precision, see Fig.\,\ref{figudv}, in the full range of $x$. This is of high interest for QCD, for example concerning the long standing issue of the high $x$ limit of the $d/u$ ratio. For the LHC searches resolving these huge uncertainties is crucial as the luminosity upgraded LHC will explore the high mass Drell Yan regions which currently are masked by the PDF uncertainties, see Sect.\,\ref{secemplhc}. 

 One should note that the distributions of the up and down quarks are related, via sum rules, to those of the heavier quarks. This has been demonstrated recently by the ATLAS Collaboration which discovered~\cite{Aad:2011dm} that there should exist a light flavour democracy
including the strange quark: finding that the ratio $(s+\overline{s})/(\overline{u}+\overline{d})$ was about $1$ and not suppressed, it was derived~\cite{Aad:2012sb} that the  light sea, 
$x\Sigma= 2x(\overline{u}+\overline{d}+\overline{s})$ was enhanced by $8$\,\%. While
ATLAS has  derived this result from essentially a moment measurement of $xs$, 
the LHeC will determine the strange quark distribution from charm tagging in CC directly and as a function of $x$ and $Q^2$ over several orders of magnitude as has been demonstrated in the CDR. This is an example of what the difference is between $pp$,
the LHC which can still lead to valuable constraints and insight for PDFs, and $ep$, the LHeC which is the ultimate machine to  measure them. Too often that is misunderstood and hopes are expressed one could determine PDFs from $pp$. This is not the case in terms of principles and precision, and it has the philosophical problem, that new physics, hopefully still appearing, as for example a contact interaction effect in $pp$, shall not be misinterpreted as a PDF effect. Two unknowns, PDFs and new physics, require two independent sources of input.

\begin{figure}[h]
\vspace*{.03cm}
\centerline{\includegraphics*[width=75mm]{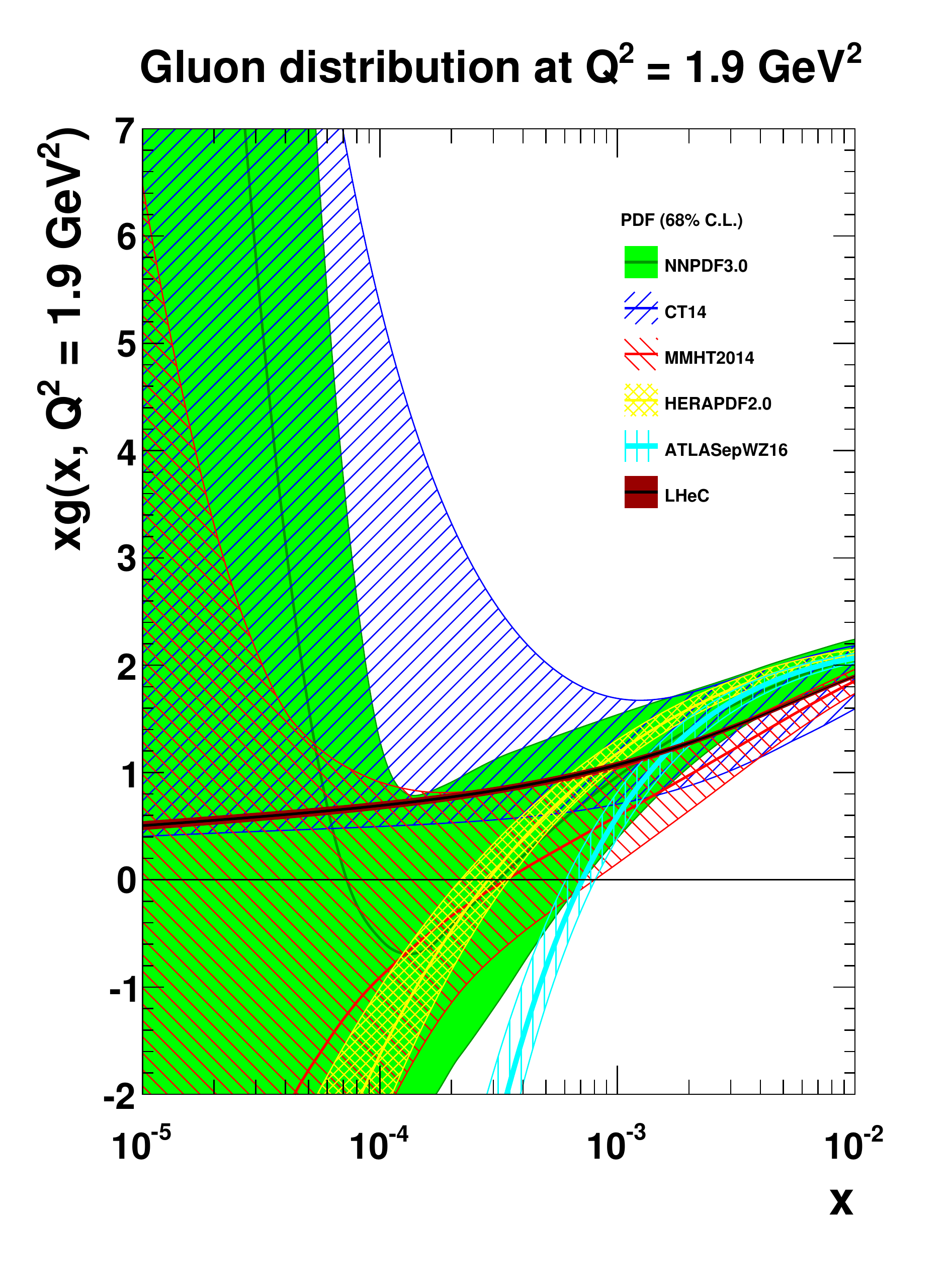}
\includegraphics*[width=75mm]{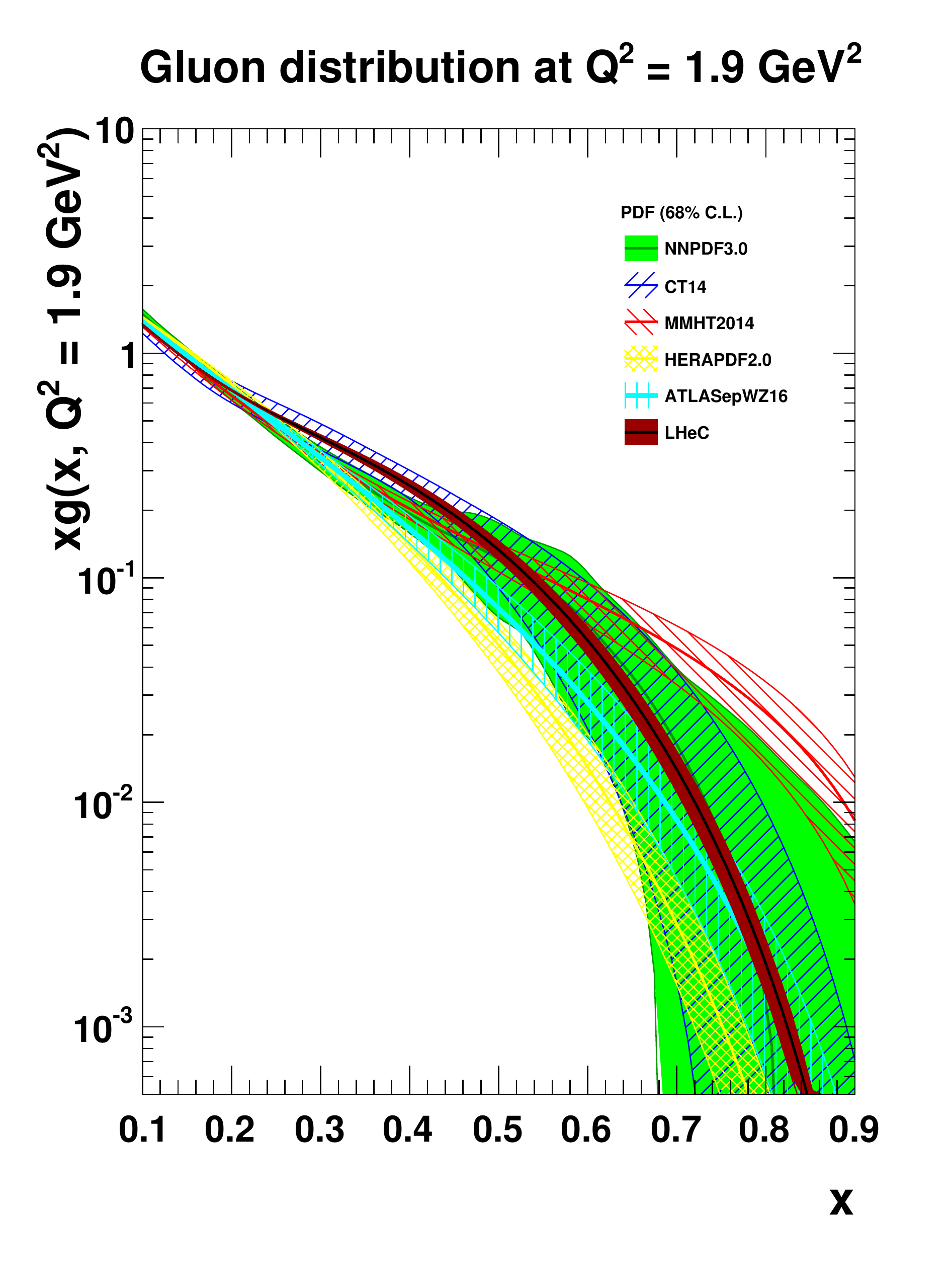}}
\vspace*{-.5cm}
\caption{\footnotesize{Determination of the gluon momentum  distribution in the proton. The expected total experimental uncertainty on $xg$ from the LHeC (dark purple bands) is compared with the most recent global PDF determinations which include the final HERA data, covering for $xg$ a range from $x\simeq 5~10^{-4}$ to $x \simeq 0.6$, and much of the LHC data from Run I. Left: $xg$ at small $x$; Right at large $x$.}}
\label{figxg}
\end{figure}
Another important example for the singular PDF prospects with the LHeC regards the gluon distribution. That is of fundamental importance because, as we discovered with HERA, the proton's momentum below $x \simeq 0.01$ is almost completely occupied by gluons, $xg$ is much larger than any for the quark distributions, in the DIS region of $Q^2$ exceeding $M_p^2$ and small $x$. Moreover, the dominant Higgs production channel in $pp$ is gluon-gluon to Higgs fusion, via top loops. The intent to measure the Higgs properties as precisely as possible at the LHC requires to control $xg$ in the region of $x$ between $0.001$ and $0.1$ to percent precision which is not achieved. The large $x$ behaviour of the gluon distribution is particularly uncertain but 
especially relevant for testing QCD, as with the factorisation theorem discussed above, and for controlling the production of new states as in gluino pair production, see Sect.\,\ref{secemplhc}.  For reliably measuring $xg(x,Q^2)$ on needs to accurately measure the $Q^2$ derivative of the DIS cross section, which, not surprisingly, came out as the only  way to accurately access $xg$ at HERA, not diffraction, not $J/\psi$ production, neither open charm which are all theoretically less clean and experimentally not competitive with inclusive DIS.

The current status on the determination of $xg$ is illustrated in Fig.\,\ref{figxg}.
A HERA physicist recognises the limits: the determination of $xg$ from $\partial F_2/\partial \ln Q^2$ at HERA reaches not below $x_{min} \simeq 2 \cdot 10/(0.7 s) = 3 \cdot 10^{-4}$ where the factor of $2$ is due to the splitting function, $10$ is the minimum $Q^2$, $0.7$ the maximum $y$ and $s=10^5$\,GeV$^2$. There arises consequently a huge uncertainty of $xg$ below such $x_{min}$ as is obvious from Fig.\,\ref{figxg} (left). Practically there is a very large uncertainty for $x$ below $0.001$. We therefore had not been able to establish the gluon density and neither small $x$ theory subtleties as resummation or BFKL effects at HERA. There are attempts to constrain $xg$ at lower $x$ through heavy flavour production at the LHC. The real determination of the gluon density at small $x$ in the appropriate theory context has to wait for the first LHeC data. The prospect is seen in Fig.\,\ref{figxg}. There, one also recognizes the HERA limits on the $xg$ determination at large $x$. The gluon density at $x=0.7$ is uncertain by two orders of magnitude, it could be equal to $0.001$ but also equal $0.01$ or $0.1$. The prospects to derive $xg$ with the LHeC are outstanding, numerically one finds a total (statistical and systematic) uncertainty, at $Q^2 =1.9~(M_Z^2)$ of $5.3~(0.2), 0.6~(0.2)$ and $21~(2)$\,\% at $x= 5 \cdot 10^{-5}, 0.01$ and $0.7$, respectively. The $Q^2$ evolution helps in reducing the uncertainty, it yet needs to be verified at the corresponding level for which, at small $x$ one will need the complementary access on $xg$ via $F_L$, see the Appendix.

\subsection{DIS and Confinement}
The LHeC nowadays is often reduced to a PDF delivery machine or a Higgs facility. While these may currently indeed be its most striking characteristics, there is a wealth of far reaching, genuine DIS physics ahead with the LHeC. QCD may break, become embedded in a higher symmetry, odderons and instantons await to be discovered. Many fundamental questions are not answered while just been raised as with HERA.  That regards, for example, the structure of the deuteron, neutron, pion, nuclei, photon and that of the diffractive exchange often called Pomeron. It comprises the many ways to modify and extend the collinear picture of parton distributions such as the transverse momentum dependent unintegrated parton distributions, the amplitude based generalised parton distributions or the non-DGLAP evolving partons. Jets and jet's substructure, characteristics of quark and gluon jets, vector mesons - a huge field of unexplored physics is awaiting an LHeC experiment. The top quark will appear light when $Q^2$ exceeds $M_t^2$ which holds for the LHeC.  The theory of heavy flavour production in $ep$ is not tested well enough, such that fixed flavour or variable flavour number schemes coexist which causes important principal uncertainty in QCD and practical uncertainty for predictions for LHC processes. DIS at high energies is a means to precisely test the electroweak theory in the space-like region with unique measurements as of the scale dependence of the weak mixing angle from below the $Z$ mass up to $1$\,TeV, or of the weak neutral current couplings of light quarks. The LHeC therefore will open a new laboratory for deep inelastic physics of hitherto unseen importance, being well comparable to the present GPD LHC experiments which it is designed to complement. 
Two examples may illustrate this rich potential.

\begin{figure}[h]
\vspace*{.3cm}
\centerline{\includegraphics*[width=90mm]{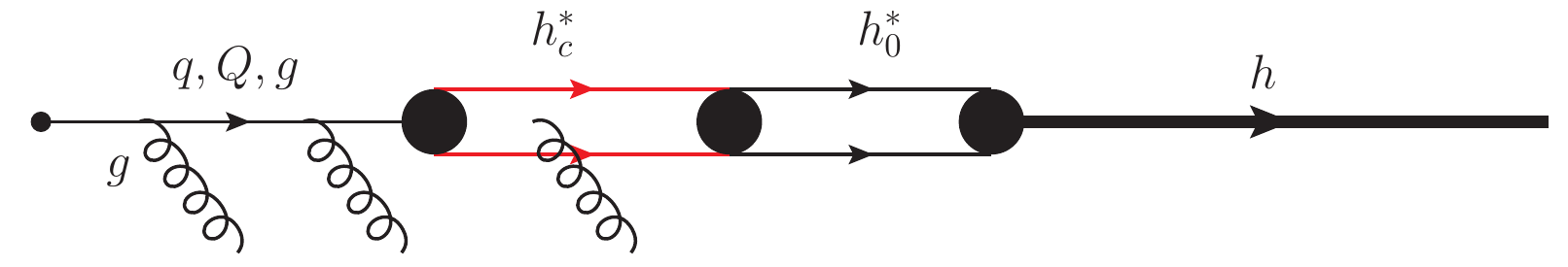}}
\caption{\footnotesize{Time evolution of hadronisation from a parton created in an interaction on to the formation of coloured and white states preceding the emission of a hadron, from~\cite{Accardi:2009qv}.}}
\label{figtube}
\end{figure}

QCD, as much as it seems to work for the description of high energy interactions as at the LHC, still has a very fundamental problem that is the explanation of the confinement of partons inside the proton. This caused Jaffe and Witten, in 2000, to propose including 
Quantum Yang-Mills Theory as one of the seven millenium prize questions, in which 
they indicated that QCD indeed shall be a consistent non-perturbative Quantum Field Theory which has to have three features: i) a mass gap between confined massless gluons and their massive bound states, ii) confinement of partons compatible with SU(3) invariant free hadrons and iii) chiral symmetry breaking to incorporate current algebra~\cite{Jaffe:2000ne}. Thus there is much more behind the strong interaction theory than is naively known. Deep inelastic scattering is ideally suited to approach the confinement puzzle by i) studying the stunning phenomenon of DIS diffraction where the proton remains intact despite a hugely energetic collision, and ii)  by exploring the process of hadronisation, i.e. the propagation of partons in spacetime, their interaction and the coalescence into colour-singlet states. The role the LHeC can play in these investigations has been discussed in the CDR, see also~\cite{Accardi:2009qv}. The LHeC allows to follow the particle formation, sketched in Fig.\,\ref{figtube}, over a long distance which can be prescribed with the scattering kinematics.
Measurements of Deeply Virtual Compton Scattering (DVCS) processes can establish
a dynamic picture of the 3D structure of the proton as is entailed in Generalised Parton Distributions. In this area, the polarised-polarised $eh$ scattering configuration at low energies provided by a future EIC and the high energy LHeC microscopy may indeed well complement each other. Important further insight is gained with the study of the hadronisation process in protons and nuclei, especially with the variation of the energy transfer $\nu=E_e-E'$.

\section{Empowering Physics at the LHC}
\label{secemplhc}
The LHC has currently delivered $100$\,fb$^{-1}$ of integrated luminosity to ATLAS
and CMS. In 2012 both discovered the Higgs boson, with about $10$\,fb$^{-1}$, but since
then, despite many expectations and predictions no further particle or symmetry
beyond the SM. LHC physics therefore focusses on two avenues, both requiring a
substantial increase of the luminosity: i) searches for new phenomena at the 
edges of phase space, i.e. states of high mass $M \propto \sqrt{s x_1 x_2}$ initiated by the
interaction of two partons carrying momentum fractions $x_{1,2}$ of their parent protons,
and ii) measurements of maximum precision to test the SM at a new level. The addition
of the LHeC is the ideal complement for both tasks, and it would be crucial also in the 
interpretation of new physics should that appear. Empowering the physics at the LHC
through $ep$ is related to the ultimate precision and clarity with which PDFs 
and the strong coupling constant will be determined at the LHeC as is discussed in this article.

\begin{figure}[h]
\vspace*{.3cm}
\centerline{\includegraphics*[width=91mm]{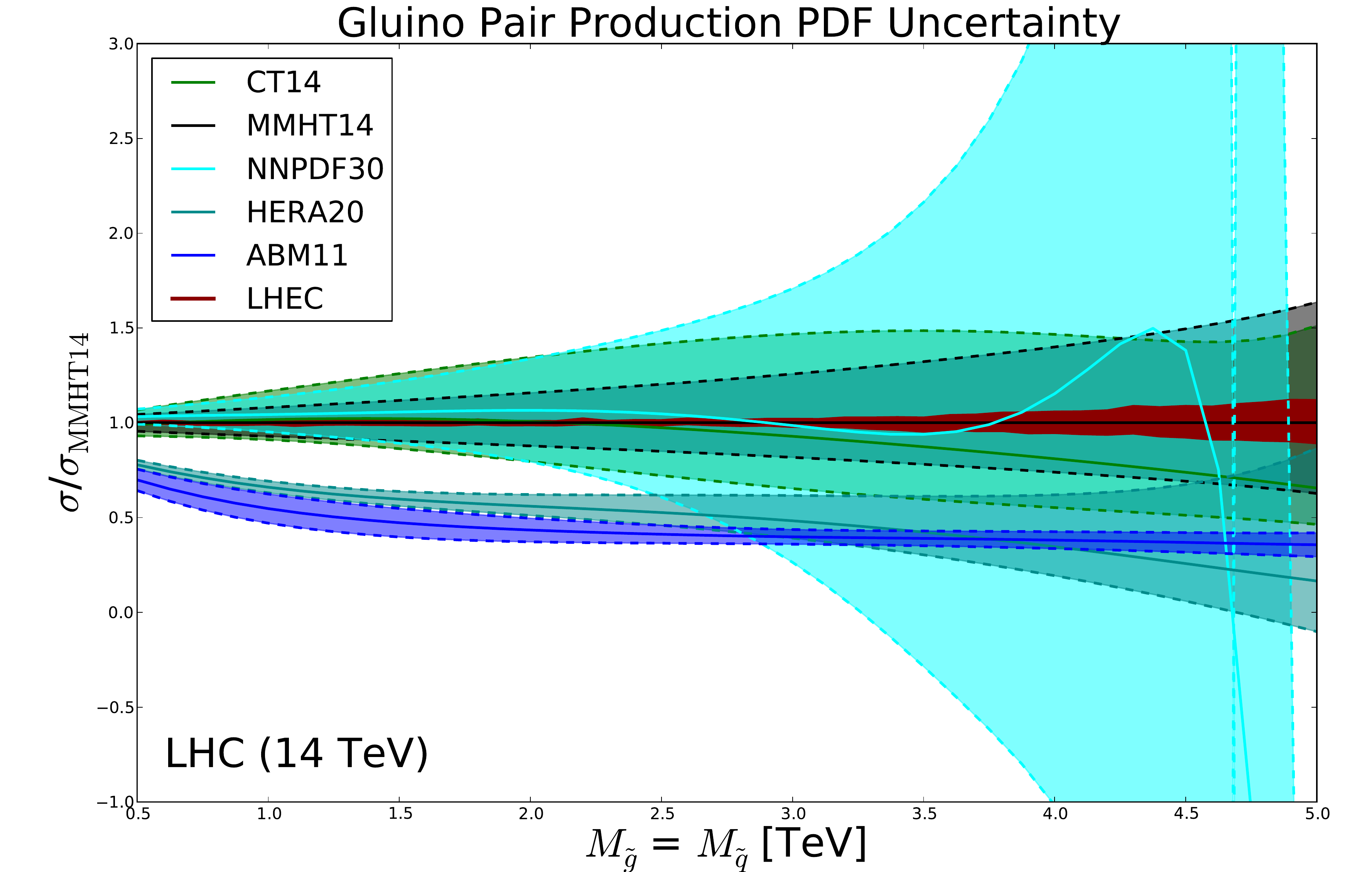}
\includegraphics*[width=88mm]{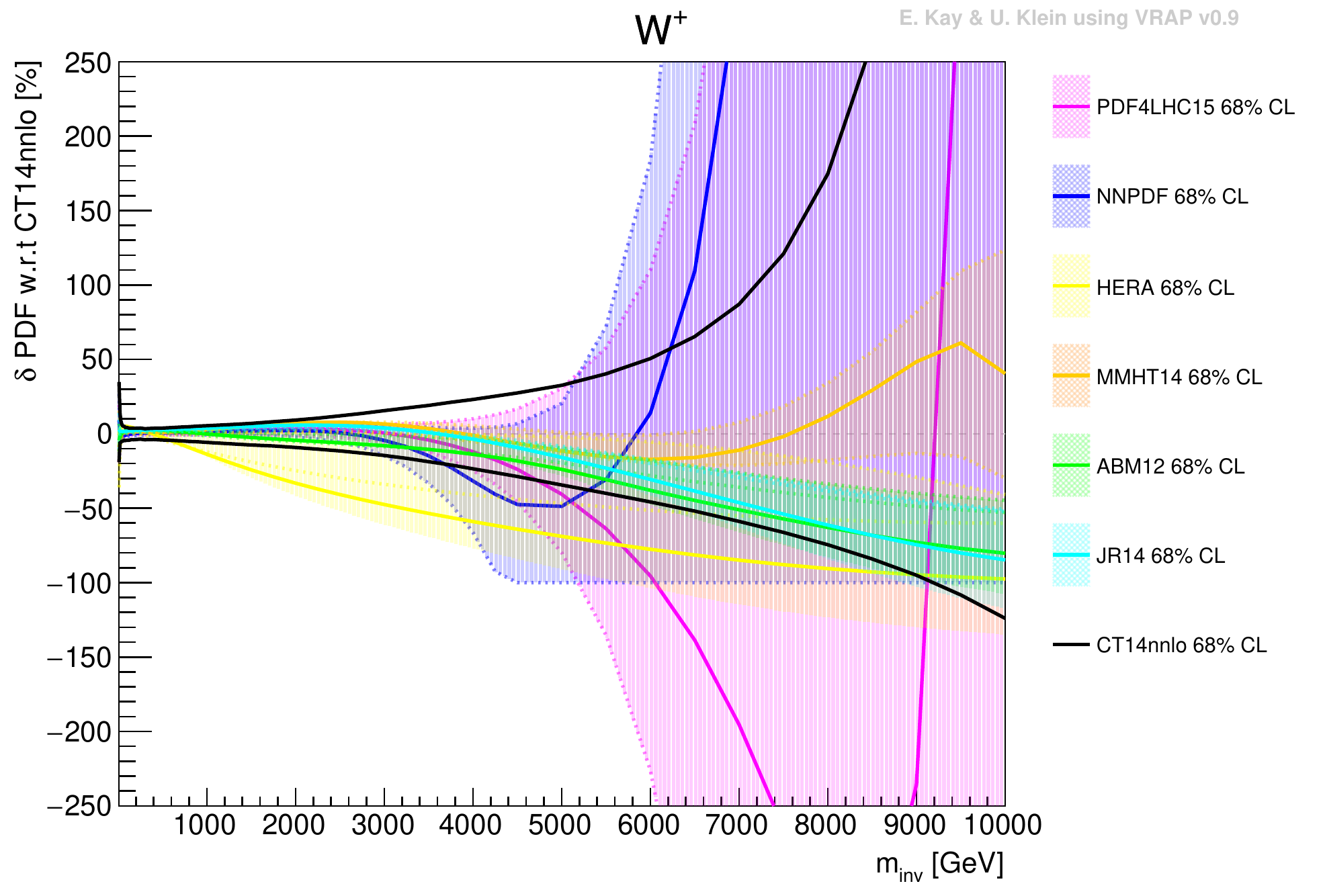}}
\caption{\footnotesize{Uncertainties of LHC production cross sections from recent
determinations of parton distributions. Left: Gluino pair production as functions
of the gluino mass. The inner red band
represents the projected LHeC PDF uncertainty~\cite{borkra}. Right:
High mass $W^+$ production~\cite{kaykle}.}}
\label{figgw}
\end{figure}

{\it{The physics motivation}}, as Guido taught us, {\it{for the LHeC is tightly related to the future continuation of the 
hadron collider experiments, HL LHC and FCC}}~\cite{ga15}.  
With increasing luminosity the LHC accesses larger and larger masses $M$. A severe problem into which it runs is that the discovery reach is thereby moved into the region
of large Bjorken $x$ where  the gluon as well as the valence and sea quark distributions are basically unknown. This is illustrated in Fig.\,\ref{figgw}. The left side shows the uncertainty of a prediction of gluino pair production in the MSSM
$gg \rightarrow \tilde{g}\tilde{g}$. At the HL LHC one may reach values up to $4$\,TeV corresponding to an average $\sqrt{x_1x_2}$ of about $0.5$, or even higher depending
on couplings to new physics. One observes how the uncertainty  rises with $M$ which
degrades the limit  values one may wish to set and complicated an interpretation
should there be new physics appearing. The right side shows the uncertainty of the
prediction for heavy $W$ bosons as are expected in grand unified theories. The uncertainty,
of the quark initiated process, rises very strongly towards large masses.
This effectively acts like a barrier for the high luminosity upgrade: huge efforts
are made to access high masses, but the proton structure is too uncertain for safely
exploring the new and most promising area of phase space. The PDF uncertainty
is already now the largest one in such searches which currently set 
limits of $5.1$\,TeV, obtained for sequential coupling and ATLAS data of $36.1$\,fb$^{-1}$~\cite{Aaboud:2017efa}. The luminosity upgrade necessitates an electron beam upgrade
for exploiting its real potential. {\it{We often hear the statement that all the relevant
information on PDFs can directly be obtained from the LHC without need of the LHeC, 
that is not really true, certainly not at the same level of precision}}~\cite{ga15}.       

The second avenue of LHC physics, more and more recognised, is ``precision", that is the desire and need to test the Standard Model as severely as possible. This concerns primarily the physics of the Higgs boson because its characteristics are not well determined and it may be the portal to new physics, see below. Moreover, the LHC is surprisingly able to perform
certain stringent tests of the electroweak theory, with the recent ATLAS measurement of the
mass of the $W$ boson~\cite{Aaboud:2017svj} to $19$\,MeV precision as a most striking 
example. In this result, the PDFs are the single largest systematic uncertainty,
of $9.2$\,MeV, resulting from quite a magic cross correlation analysis, which works through
the constraint DIS data impose on the total light quark sea, and by considering
only a restricted set of PDFs. The LHeC is estimated to predict this $M_W$
measurement to $2.8$\,MeV precision as derived from the uncertainties of the LHeC PDF 
set~\cite{stefanomw} which eliminated the PDF problem in the $M_W$ mass measurement at the 
LHC. It is moreover interesting to note, that  $M_W$ can also be extracted
from the LHeC data as a fundamental parameter of the electroweak theory, with an estimated
total uncertainty of [$14$\,(exp) and $10$\,(pdf)]\,MeV~\cite{danielmw}. This is an example of the ultra high precision the LHeC will provide in the development of the SM and its possible eventual failure. Many other examples could be discussed here too, such as the scale dependence of the weak mixing angle, the determination of the $V_{cs}$ and $V_{tb}$ CKM matrix elements or the per mille accurate measurement of $\alpha_s$  (cf Appendix).
\section{The LHC as a Precision Higgs Physics Facility}
{\it{It is crucial for future experiments at the LHC and elsewhere
to confirm the properties of the Higgs and the absence of new physics}}~\cite{Altarelli:2014roa}.
The Higgs mechanism generates the mass of the weak gauge bosons and the elementary fermions. Its 
simplest realisation contains a scalar field, $J^{CP}=0^{++}$, which is referred to as the Higgs boson.
Following the discovery of this particle, in 2012, much effort has been spent by ATLAS and CMS to pin
down its quantum numbers and couplings to its decay products which are expected to depend linearly
on their mass squared. The mass of the Higgs boson of $125$\,GeV implies that a particularly large number
of channels is open dominated by the $H \rightarrow b\overline{b}$ decay with a relative probability of
$58$\,\% in the SM. Due to the fundamental role of the Higgs phenomenon, and in the absence of other
new physics, its exploration has moved into the centre of attention at the LHC and 
for planning the future of collider particle physics. 

\begin{figure}[h]
\vspace*{-.2cm}
\centerline{\includegraphics*[width=115mm]{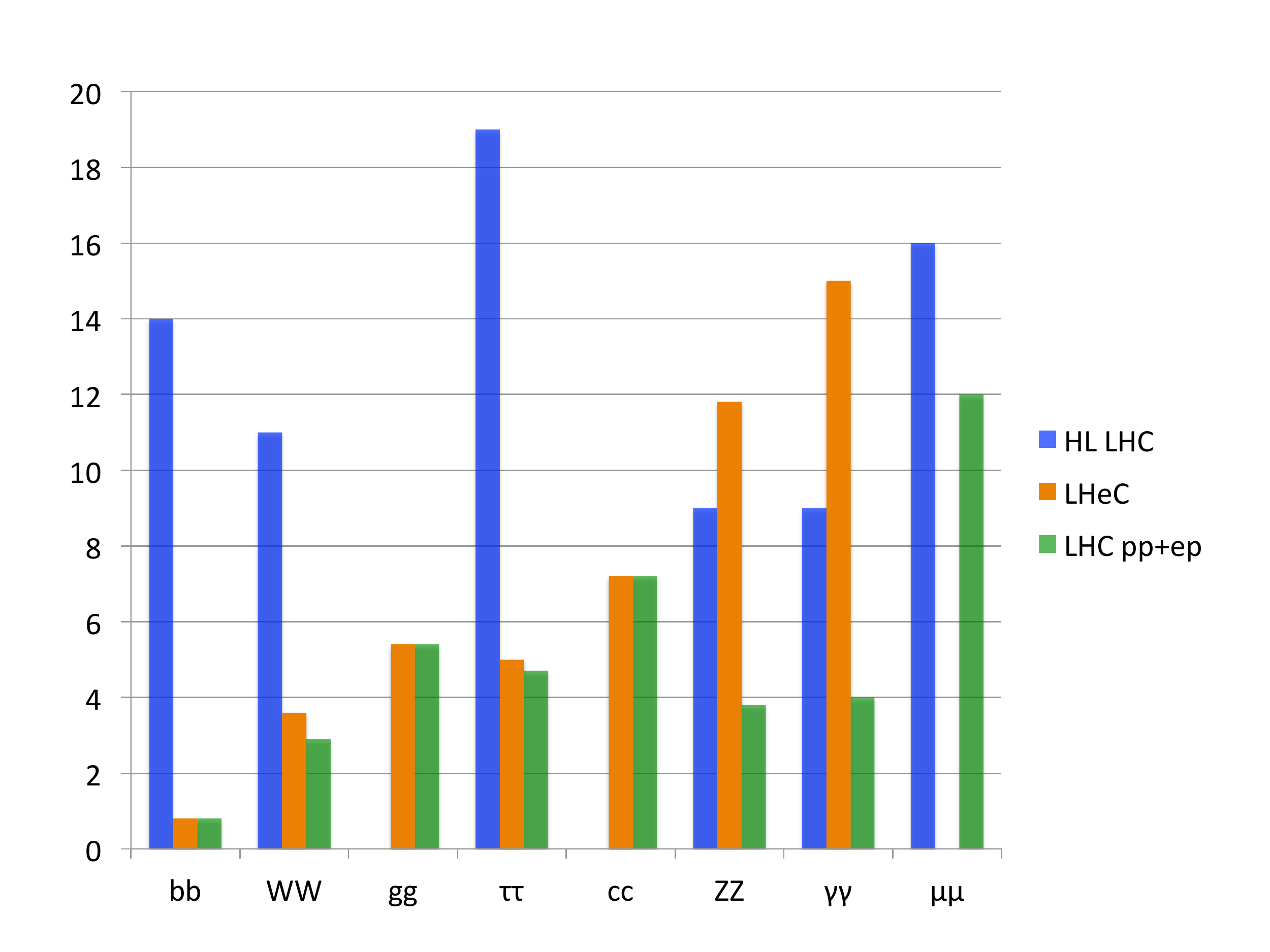}}
\caption{\footnotesize{Preliminary estimates of the expected
 determinations of the relative signal strength $\delta \mu / \mu$, in \%, of the
Higgs decay to various particles
at HL LHC. Blue: the ATLAS expectation for the HL LHC~\cite{AtlasPub14}; 
Brown: LHeC prospect~\cite{muk18}; Green:
LHC prospect for an $ep/pp$ combination, see text. The decay channels are arranged in descending branching
ratio order, from $57.7$\,\% for $H \rightarrow b\overline{b}$ to $0.02$\,\% for $H \rightarrow \mu \mu$. The improvement for all channels wrt. the HL LHC is due to new information on various decays from the LHeC
and to an essential removal of the QCD uncertainties with the provision of N$^3$LO precision PDF and
$\alpha_s$ measurements by the LHeC. An empty column, such as for the LHeC (brown) 
in the $\mu \mu$ channel, indicates that this may not be accessible to notable precision.}}
\label{fighcoup}
\end{figure}
In this context the first question concerns the potential of the HL LHC
for probing the Higgs properties. The LHC is the only existing Higgs facility and for very long time it will be the
only one with reliable access to very rare decays such as $H \rightarrow \mu \mu$. At the same
time it has considerable difficulties to accurately measure certain abundant decays, for which one needs
to employ the associate vector boson production, which just recently allowed the
$H \rightarrow b\overline{b}$ decay to be recognised~\cite{Sirunyan:2017elk,Aaboud:2017xsd}.
There are, moreover, considerable theoretical uncertainties, partially related to the insufficient
control of $xg$, $\alpha_s$ and non-availability of genuine N$^3$LO PDFs
which can be cured with the LHeC. It has been argued
that the LHC can be ``turned into a precision Higgs facility"~\cite{utapH} with the addition of the
LHeC which has no pile-up and a clean final state signature which distinguishes
$WW \rightarrow H$ from $ZZ \rightarrow H$ production. 
Considerable work is ongoing to study this at the required level, and initial results
and the direction of this exploration are illustrated here. In addition, as sketched in 
Sect.\,\ref{secnewphys} below, there are unique opportunities  with the LHeC 
to access new Higgs physics, in the production (as of charged Higgs bosons) and 
decay (as into dark matter), which would lead beyond the SM.

 A particular question is whether the total decay of the Higgs
boson is exhausted by its branchings into the known particles, such as pairs of 
$b,~W,~g,~\tau,~c,~Z,~\gamma$ and $\mu$ and a few mixed combinations such as $Z \gamma$.
Fig.\,\ref{fighcoup} illustrates the expected precision of the measurement of the signal strength $\mu$
for the eight most abundant decay modes in the SM. The plot shows the total uncertainty, in blue,
as estimated by ATLAS~\cite{AtlasPub14} for the HL LHC, i.e. $3$\,ab$^{-1}$ of integrated luminosity.
A second column shows the expected precision on $\mu$ for the LHeC for an assumed luminosity
of $1$\,ab$^{-1}$ collected in electron-proton scattering. The results refer to a charged 
current interaction where the Higgs boson is emitted from a $W$ exchanged in the
$t$ channel.  
The third, green column illustrates the expected accuracy for the combined $pp$ and $ep$ result at the LHC.
 It is to be noted for the LHeC that so far
only the $b\overline{b}$ and $c\overline{c}$ results for the LHeC are due to complete 
simulations~\cite{utabc} while the other numbers are still of tentative nature~\footnote{
The signal strength result for  
the prominent $H \rightarrow  b\overline{b}$ decay  is $0.8$\,\%, or a $0.4$\,\% for the 
coupling determination. This is a most stunning precision and used to optimise the LHeC
detector design, especially for forward $b$ tagging acceptance and optimum resolution
of the hadronic calorimeter and joint track-cluster reconstruction. The $bb$ decay
was initially studied for $ep$
when the physics of LEP\,x\,LHC was explored~\cite{Grindhammer:1990vs}. Looking back at 
that analysis one recognises the impressive progress in the experimental and multi-variant
analysis techniques and values the $10^{34}$ luminosity prospect particularly high.
The $c\overline{c}$ decay is considered to be not accessible  at the LHC while the preliminary $ep$ 
result~\cite{utabc}  is $7.2$\,\% which transforms to a $3-4$\,\% precise determination of the Higgs coupling to the charm quark.}.
For the LHC itself one would expect also improvements related to progress in the analysis techniques and a
possible final combination of the two experiment results. 

Based on the numbers presented in Fig.\,\ref{fighcoup} one may calculate 
the sum $B_H$ of the Higgs boson branching fractions, $br_i,~i=bb,WW,...$,  
and their total uncertainty. The LHC result, assuming 
complete independence of the $br_i$ measurements,  comes out to be
$B_H(pp) = 0.89 \pm 0.12$. The LHeC, assuming it indeed established the $H \rightarrow gg$ decay,
which has a branching fraction of $8.6$\,\%,  to better than $10$\,\% precision, would reconstruct
this sum as $ B_H (ep) = 1.00 \pm 0.02$. The combination of $ep$ and $pp$, with the 
additional reduction of the QCD uncertainty in $pp$ through information from $ep$, can be expected to accurately determine
the total sum as
\begin{equation}
 B_H (pp+ep) = \Sigma_i~br_i = 1.00 \pm 0.01.
\label{eqBH}
\end{equation}
This prospective result  means that the LHC  can indeed be turned into a a very powerful
Higgs facility through the addition of $ep$. This is possible at moderate cost compared 
to other new facilities currently considered to which $pp/ep$ would be
the ideal complement. The precise measurements on
$b\overline{b}$ and $c\overline{c}$, the identification of the $gg$ and other decays, still
under study,  the reduction of 
QCD uncertainties for the $pp$ results, as well as the experimental and
the theoretical cleanliness of the $ep$ Higgs production process~\cite{Blumlein:1992eh} 
are crucial improvements the LHeC entails. 
 The result of Eq.\,\ref{eqBH} would narrow
any window for exotic Higgs decays, which currently is wide open (to about $30$\,\%),
to a per cent and therefore represent one of the most crucial new tests of the SM. 
Failure to reconstruct $B_H$ to $1$ would, on the contrary, be an ultimate hint to
the existence of new physics.  Further
interpretations of that experimental input on $\delta \mu_i/ \mu_i$ 
may be made in joint fits of the couplings as are frequently
done in prospective Higgs studies, such as in the recent EFT based analysis for
the Linear Collider prospects~\cite{Barklow:2017suo}. Extra input in $ep$ is gained from neutral
current measurements, especially of the $ZZ \rightarrow H \rightarrow b\overline{b}$ reaction.

 It should be seen that a SM independent
measurement of the sum of  branching ratios $B_H$ would require to access the total width
$\Gamma$. That is considered to be an important advantage of $e^+e^-$ colliders.
It is intended to study this for the LHeC. The neutral current reaction $e^-p \rightarrow e^- H X$
can be kinematically fully constrained. The sizeable total NC Higgs cross section corresponds to
$25k$ events which ought to be useful for a yet challenging measurement of $\Gamma$. 
One should further note that only the LHC (and its possible extensions HE LHC and FCC), 
neither the LHeC nor a future $e^+e^-$ collider,
has the possibility to access rare Higgs decay channels. The total $H$ production cross section at the LHeC,
very similarly for the ILC or CepC, is $200$\,fb$^{-1}$. For a luminosity of $1$\,ab$^{-1}$ this
provides $40$ $H \rightarrow \mu\mu$ events only. It so appears that the  combination of
$pp$ and $ep$ at the LHC form a unique, very powerful facility to lead Higgs physics in the thirties
to an unprecedented art even if some of the here assumed prospects may alter in the course 
of the ongoing further study.

\section{New Physics at the LHeC}
\label{secnewphys}
Compared to the LHC proton-proton interactions, the LHeC is a cleaner configuration because not only it has no pile-up of events but it has a unique NC or CC final state free of colour connection effects. Compared to the ILC and CepC, with perhaps initially $\sqrt{s}=0.25$\,TeV energy,  it has 
an about five times higher centre of mass energy. The $ep$ configuration would fuse electrons and partons to leptoquarks, but it primarily is viewed as a space-like probe, with photons and $Z$ bosons in neutral currents and $W^{\pm}$ in charged currents, of the substructure and dynamics of matter. In these interactions, however, new, heavy particles, such as the Higgs boson, may be produced and new phenomena beyond the Standard Model be revealed. With its  luminosity prospect~\cite{Bruening:2013bga,fccnote} of O($10^{34}$)\,cm$^{-2}$s$^{-1}$ the LHeC has a huge potential for new physics to be discovered in $ep$. One should note that such a luminosity allows the 15 years of HERA data taking
to actually be pursued in just one day. None of the two $e^+e^-$ colliders mentioned,
which currently have certain probability of being realised, promises
to have significantly higher luminosity. The HL LHC, with $pp$ and $ep$, and a new $e^+e^-$
machine set to $ZH$ production would be a strong triple for the next
phase of energy frontier collider physics, in the third decade of our century. This would substantially strengthen
CERN's leading role in particle physics while a healthy competition could
rise in Asia, prior to the realisation of a next, more demanding step
 with raising the energy for $pp$ collisions at CERN.

 With increasing intensity and novel approaches, the genuine BSM potential of the LHeC is now being studied, in often comparative analyses, and it is
generally observed that the BSM potential in $ep$ is much stronger than  thought to be hitherto. A new generation of theorists challenges the conventional believes. It is regretful  we cannot discuss this with Guido Altarelli who had put to question whether $ep$ could indeed be a competition with $e^+e^-$ or hadron colliders by calling it, in contrast to those,  {\it{not~a~main~discovery~machine}}~\cite{ga15}. A few examples are here given subsequently which illustrate new avenues and results towards a new exploration of the BSM physics potential in $ep$ which leads further than the CDR, because of the challenges set with the absence of obvious new phenomena at the LHC and because apparently only slowly the LHeC prospect gets recognised.

\begin{figure}[h]
\vspace*{.3cm}
\centerline{\includegraphics*[width=110mm]{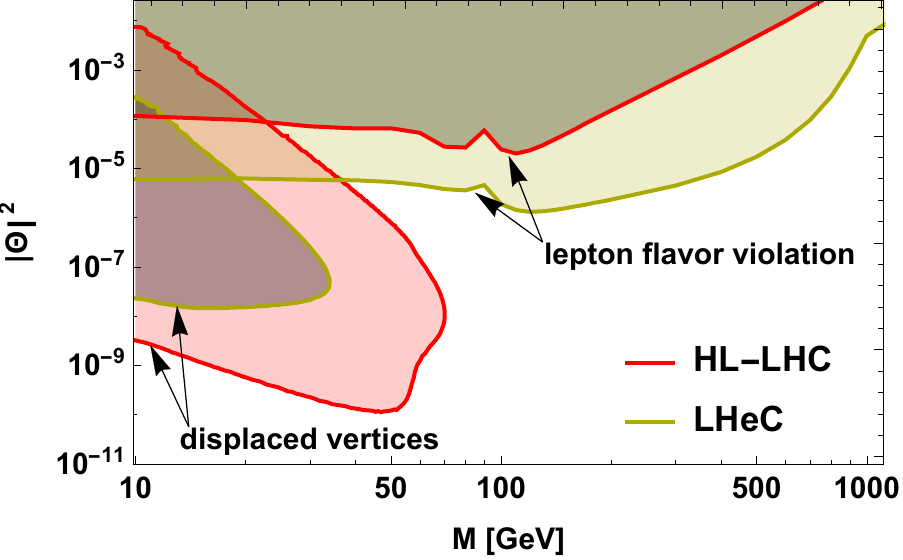}}
\caption{\footnotesize{Sensitivity of the HL LHC and the LHeC to the existence of sterile neutrinos in the plane of the active sterile mixing angle $\Theta$ and the heavy neutrino mass $M$. The left, lower mass sensitivity is due to displaced vertex sensitivity while the large mass sensitivity is due to lepton flavour violating multi-jet signatures~\cite{Antusch:2016ejd}.}}
\label{figSN}
\end{figure}
A central question in particle physics is that of the origin of mass. The LHeC provides fundamental insight to the origin of baryonic mass, with its far reaching exploration of gluon dynamics in hadrons, of dark mass, through Higgs-to-invisible measurement prospects, of boson and top mass, through unique contributions to their determination  and to the Higgs mass, through the approach to the Higgs self-coupling~\cite{Kumar:2015kca}. A recent study~\cite{Antusch:2016ejd} draws the attention to the origin of neutrino masses which may require an extension of the Higgs sector 
or the existence of neutral fermions. Such heavy neutrino mass eigenstates, often termed sterile neutrinos, through their interaction with weak bosons, give rise to a rich phenomenology for their possible different production and decay modes, the dependence on mixing parameters or the occurrence of lepton number or lepton flavour violation effects to which the three collider configurations exhibit different sensitivity.
 Fig.\,\ref{figSN} displays the result of a comparative study, here focussed on HL LHC and LHeC, of the sensitivity of future collider configurations to the existence of heavy neutrinos, which in $ep$ are primarily produced in the CC reaction $ep \rightarrow NX$. The LHeC reaches a sensitivity of  neutrino mass up to $\sim 1$\,TeV at mixing angles squared down to about $10^{-3}$ extending the $\Theta^2$ range of the HL LHC 
by an order  of magnitude to smaller values, owing to the cleanliness of the final state.
For  lighter masses, up to about $30$\,GeV, very small mixing angles 
of $\Theta^2$ down to $2 \cdot 10^{-8}$ are reached. In this area, from an initial study, the HL LHC sensitivity looks stronger than that in $ep$, see Fig.\,\ref{figSN}. 

\begin{figure}[h]
\vspace*{.3cm}
\centerline{\includegraphics*[width=110mm]{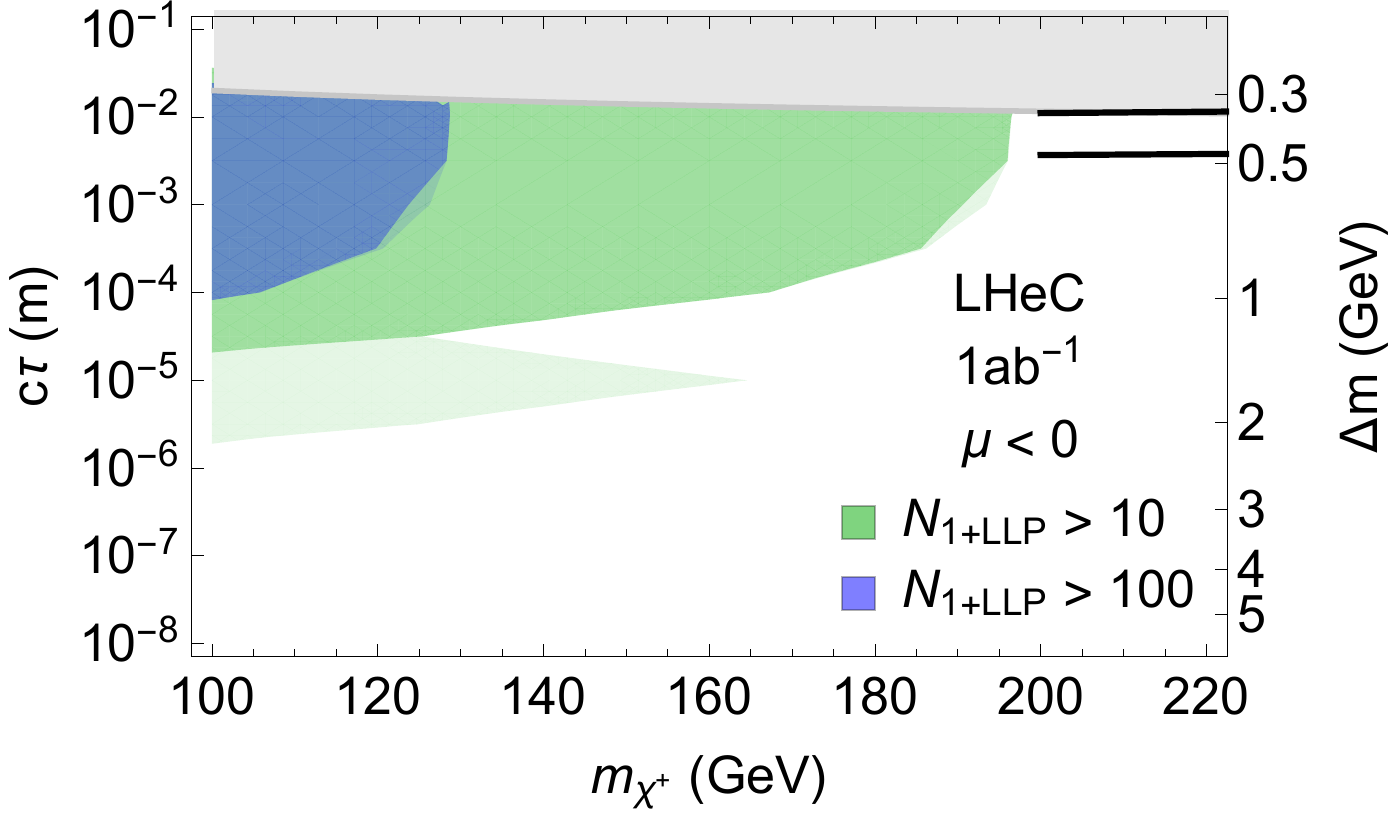}}
\caption{\footnotesize{Regions in the mass-lifetime Higgsino parameter plane
accessible for LHeC events with at least one LLP. Black curves are projected bounds from
searches at the HL LHC, from~\cite{Curtin:2017bxr}.}}
\label{fighino}
\end{figure}

Another  example for the unique potential to discover new physics at the LHeC has just been presented
with a study of the search for long lived Higgsinos~\cite{Curtin:2017bxr}. These occur in the MSSM together
with Wino and Bino states. Their mixing generates 4 neutralino and 2 chargino mass eigenstates. It has
been pointed out that while such particles may be detected at hadron colliders for large mixing,
a particularly clean final state is required to discover Higgsinos for small mixing which leads
to mass degeneration. The result of quite a thorough analysis of a t-channel $VV \rightarrow \chi \chi$
pair production of supersymmetric particles $\chi$ from exchanged vector bosons $V$ is reproduced
in Fig.\,\ref{fighino} which displays the number of observed events with at least one long lived particle (LLP) 
as a function of the chargino mass and lifetime. It can be seen that the LHeC is much more sensitive 
at small $c \tau$ than the LHC. This advantage holds similarly for exotic Higgs decays which can occur with sizeable
branching fractions as is argued in~\cite{Curtin:2017bxr} too. 

A number of further  analyses have been published or are being pursued, as are
summarised in \cite{utabc}, which demonstrate
a very high and often unique potential opened by the LHeC for the discovery of exotic Higgs physics. These
regard, for example, the exotic Higgs decays, as into dark matter with a branching sensitivity to $5$\,\%, or  
into two new scalars~\cite{Liu:2016ahc} which provides sensitivities which surpass the 
HL LHC very much. There are also LHeC related papers published on the top-Higgs coupling~\cite{Coleppa:2017rgb}
on the possible discovery of charged Higgs bosons~\cite{Azuelos:2017dqw} and others. 
These and further ongoing   studies lead the examination of BSM $ep$ physics
much beyond the level reached with the CDR on the LHeC in 2012. 
They strongly support the conclusion reached in~\cite{Curtin:2017bxr}
that adding a high luminosity, high energy  DIS program to a $pp$ collider ``is necessary to fully exploit its discovery potential for new physics"~\cite{Curtin:2017bxr}. 

The potential to discover new physics at the LHeC reaches far beyond the Higgs sector and its exotic
extension. From very high precision measurements of the structure functions at large $x$ new
sensitivity is reached to  contact interactions into the O($100$) TeV region, subject to assumptions
on their coupling structure. An analysis of the potential to exploit the general Lagrangian structure as
embedded in EFTs has not been done yet for $ep$ even though $1053$ out of $2499$
parameters characterising deviations from the SM are related to lepton-quark 
interactions~\cite{Trott}.

While HERA had too low energies to study top quark physics, the LHeC is actually a top quark factory
with single $t$ and $t \overline{t}$ production 
cross sections of $\sim 5$ and $\sim 0.02$\,pb, respectively. This permits
high precision measurements and unique searches for new physics in many areas, such as
the anomalous $t-q-\gamma$, $t-g-Z$ and triple gauge~\cite{orhan} 
couplings, the top contribution to the proton's momentum
balance (top PDF), anomalous top-Higgs couplings~\cite{Sun:2016kek} and  CP 
nature~\cite{Coleppa:2017rgb} and others~\cite{haodis,cschwa}. 
The prospects for top $ep$ physics are a particularly strongly developing field with many
interesting results supporting the physics case of the LHeC.

New physics can certainly be expected to be discovered in the strong interaction sector. QCD is a very
rich and largely unexplored theory regarding, besides the confinement puzzle cited above,
for example i) the AdS/CFT duality, which links supersymmetric modifications of QCD to string theory in anti-de-Sitter space or ii) the instanton fluctuations of the gluon field. The LHeC may also discover odderons,
a saturation of the gluon density at small $x$,
or a breaking of factorisation, as has been discussed in detail in~\cite{stefanofac}. 
It should clarify the question why all hadron-hadron interactions of protons and ions, $AA',~pA$ and $pp$,
exhibit collective phenomena in the expected contrast to $ep$. The LHeC will push the evolution
of QCD much forward, such as the higher order (N$^3$LO and N$^4$LO) calculations in pQCD,
or the physics of non-conventional PDFs. In a list of possible QCD discoveries, presented
by C.\,Quigg with ``the unreasonable effectiveness of the SM" at the DIS workshop in 
2013~\cite{quigg}, the existence of free quarks, unconfined colour, quark
substructure and the embedding of QCD in a higher symmetric theory  are included, the discovery 
of each/any of which would be of far reaching  consequences for particle physics.  There are
strong, accumulating reasons to believe that the LHeC, as the second highest energy, high luminosity,
asymmetric collider should indeed be among the main configurations for leading beyond the
Standard Model.
\section{LHeC as an EIC}
The prime task of an electron-ion collider at high energies, besides
a detailed mapping of nuclear dynamics and structure, is to investigate the QCD dynamics 
at high parton densities. These are observed at small Bjorken $x$ and may be amplified 
in a nuclear medium.
\subsection{Low x Physics}
\label{seclowx}
{\it{.. subleading terms cannot be ignored and must also be resummed. Their effect is to push the onset 
of the truly asymptotic regime down to smaller values of x, while in an intermediate region, relevant for
 HERA data, the evolution shows a shallow dip, somewhat lower than the NLO perturbative result. The 
 resummation procedure is determined by solid guiding principles from duality, momentum conservation, 
 symmetry under gluon exchange of the BFKL kernel and accurate implementation of running coupling 
 effects}}~\cite{Altarelli:2008aj}. Deep inelastic scattering is the interaction of leptons and hadrons at 
 four-momentum squared values exceeding the mass of the proton squared, $Q^2 > M_p^2$. In this region 
 perturbative QCD may be applied as $\alpha_s(Q^2) \simeq 1/(\beta_0 \ln{(Q^2/\Lambda^2)}) < 1$, 
 where $\beta_0 = (33-2N_f)/12 \pi$ is the one-loop renormalisation group equation coefficient,
 $N_f$ the number of active flavours and $\Lambda \simeq 200$\,MeV the famous parameter which
 quantifies the value of the strong coupling, which is about $0.5$ at $Q^2 =1$\,GeV$^2$.
 Bjorken $x$ is related to $Q^2$ via $x=Q^2/sy$, where the inelasticity
 $y$ is related to the energy transfer $(E_e - E')/E_e$ and limited between $0$ and $1$. The smallest value
 of $x$ any $ep$ collider may reach in the DIS region therefore is $x_{min} = 1/s$ with $s=4E_eE_p$ in GeV$^2$.
 There is a sequence of  minimum $x$ values of $\log{x_{min}} \simeq (-2,-3,-4,-5,-6,-7)$ for fixed target
 DIS, the two US EIC versions, HERA, the LHeC and FCC-eh, respectively. One would call low $x$ the region 
 below $x \simeq 10^{-4}$ down to which the DGLAP equations have been observed at HERA to 
 approximately hold. Low $x$ physics thus is a prime
 area of research for the LHeC, and for future higher energy DIS colliders under study for CERN and China.
 
 A striking discovery at HERA was the rise of the quark and gluon distributions towards small $x$. It has been argued that
 unitarity should tame that rise through non-linear parton-parton interactions~\cite{Gribov:1984tu}, predominantly gluons because  at small $x$ the gluon density is much larger than that of sea quarks. This observation holds only for
 large $Q^2$, in excess of $\simeq 10$\,GeV$^2$, because $xg$ turns valence-like at $Q^2 \sim 1$\,GeV$^2$~\footnote{
 That observation is key because it enlarges the effective $x_{min}$ values by an order of magnitude such that
 LHeC will investigate the domain down to $x \simeq 10^{-5}$ and eRHIC to only $\simeq 10^{-3}$ which has already
 been explored with high precision at HERA.}. 
 The key theme of the physics at small $x$ is therefore the understanding of parton interaction dynamics
at high densities and for small $\alpha_s$, which is an unexplored dynamic state of matter.
 The key experimental challenge is the discovery of non-linear effects,
often  synonymously denoted as BFKL dynamics, which would end the region of validity   
of the linear DGLAP evolution equations.  This is not only a question of theoretical importance but 
of very practical relevance for the LHC and higher energy hadron colliders: the gluon cannot be directly
observed, its distribution is defined in a theoretical framework. If that changes fundamentally, the PDFs change
and the cross-section calculations alter. Note that in Drell Yan scattering of two partons a state of mass $M$
is produced with a rapidity $\eta$ related as $\sqrt{x_1 x_2 s} = M exp(\pm \eta)$. For Higgs production
the central (very forward) rapidity value corresponds to an average  $x = \sqrt{x_1 x_2}$
of $0.009~(0.0002)$. These $x$ values are lower by a factor of $7$ for the FCC. The hadron colliders
already sample very small $x$ values for the Higgs but even more so  for lighter states. The clarification 
of the low $x$ dynamics with the LHeC is therefore crucial for the HL LHC and a necessity for the
higher energy colliders. It is also of much relevance for the understanding of ultra high energy neutrino
physics. These tasks need DIS energies higher than HERA's and not lower.
   
The decisive test for BFKL dynamics in DIS proceeds through high precision measurements of the 
salient structure functions $F_2$ and $F_L$ as has been demonstrated in the LHeC CDR~\cite{Abelleira2012}.
This holds because these functions are most cleanly theoretically defined. Both, however, are needed because
if one had only $F_2$ one could attribute changes in the evolution law too easily to parametric peculiarities
of the collinear PDFs. This causes a huge challenge for a precision measurement of $F_L$ at small $x$
which is discussed in the Appendix.

The establishment of BFKL dynamics at small $x$ would be convincing only if the whole dynamics in this
range was consistently described. Consequently there are many other processes under study for the 
observation of non-linear parton interactions. One example, advocated at a recent LHeC workshop 
by Mark Strikman, is
the production of a vector meson with a rapidity gap in virtual photon-hadron scattering~\cite{mark15}.  
This reaction is viewed as small dipole elastic scattering at large $t$ and large energies, small $x$, and
may lead to an observation of BFKL dynamics. Pursued also in electron-ion scattering it will permit to study
the effect of nuclear media with regard to the regulation of shadowing and the eventually possible
observation of the black disc regime. 

As an experimentalist one admires the considerable theoretical development of low $x$ QCD, indicated also
by Guido's remarks cited above, and must admit that the information we have been providing is not enough,
HERA being too low in energy and too weak in precision and the LHC being plagued by its nature, the
colour connections in the $hh$ interaction. The LHeC, with much  less luminosity than $1$\,ab$^{-1}$, will change
that completely and has one of its major and unique tasks in the clarification of the parton dynamics at small $x$.
\subsection{Nuclear Parton Dynamics and Densities}
\label{seceA}

Electron-ion collisions provide rich perspectives for nuclear Chromodynamics. This holds for the
low energy EIC collider~\cite{Accardi:2012qut} which may reach cms energies  of $\leq 100$\,GeV as compared to
$\simeq 20$\,GeV by the CERN muon-hadron scattering experiment NMC. It holds as well for the LHeC
reaching the TeV energy scale. A difference, apart from the kinematic reach, is the dedication the EIC
may put to electron-ion collisions which at CERN would most likely remain in the shadow of
the $pp$ program. However, recent evaluations of the $ePb$ luminosity at the LHeC, presented in~\cite{fccnote},
result in an estimate of $L=7~10^{32}$\,cm$^{-2}$s$^{-1}$. One can thus collect annually about $1$\,fb$^{-1}$
of integrated luminosity which at low $Q^2$ is ample, also for semi-inclusive analyses, and implies
that a complete DIS programme in electron-ion scattering is ahead including, for example,
high $Q^2$ charged current physics and providing O($10^3$) Higgs events emerging from electron-lead
scattering.

A recent summary of novel perspectives for nuclear physics includes topics ``such as the hidden color of nuclear form factors, the relation of the nuclear force at short distances to quark interchange interactions, the effects of Òcolor transparencyÓ on the baryon-to-meson anomaly in hard heavy-ion colisions, novel exotic multiquark states, the anomalous nuclear dependence of quarkonium hadroproduction, flavour-dependent antishadowing, and the breakdown of sum rules for nuclear structure functions. I also briefly discuss the insights into hadron physics and color confinement that one obtains from light-front holography, including supersymmetric features of the hadron spectrum"~\cite{Brodsky:2017oov},
too detailed to be discussed in this contribution. It may suffice to state gratefully that Stan Brodsky has had quite
an influence on shaping the LHeC programme and supporting its built as a member of the IAC and in his 
outstanding capacity as a world leading theorist.

For the search for non-linear parton interactions one often cites the expectation, especially when 
gluon saturation is discussed in the US EIC context, that the gluon density in a nucleus $A$ is amplified
by a factor $A^{1/3}$, which may only work if the underlying proton parton density is large, i.e.
for $xg$  for $Q^2 \geq 10$\,GeV$^2$
as mentioned above. Moreover, that amplification
may be tamed by flavour dependend shadowing effects. There are, principally,
two tasks which require independent input: the exploration of  BFKL effects and the clarification of nuclear
medium and binding effects. It therefore is strongly believed that one needs higher energies than HERA's 
in order to probe for non-linear evolution in $ep$ and then to compare the result with $eA$. The high luminosity
in $eA$ at the LHeC  and the large kinematic coverage enable the derivation of the flavour decomposed light
sea quarks, the gluon and direct measurements of the strange, charm and beauty distribution in nuclei
available at the LHC.  This programme is being investigated, for a summary see~\cite{nestor}.
It will completely change the way nuclear PDFs are obtained, end the use of $\pi_0$ data, for example, and
deliver nPDFs free of any pPDF base. One then measures directly the ratio $R=f_A/A \cdot f_p$, flavour by flavour and
for $x$ between a few times $10^{-6}$ up to $1$. This is in contrast to the present method in which $R$ is
evaluated using certain set of proton PDFs and a theoretical parameterisation of nuclear effects, with reliable data
available only for $x > 0.01$. This is a most exciting prospect as has been 
outlined in~\cite{Klein:2016uxu,Helenius:2016hcu}.

The study of $eA$ collisions in the TeV regime at the LHeC would  be crucial
for the development of nuclear QCD and also for
the interpretation of the data coming from the heavy-ion programme. Its
importance would be threefold~\cite{nestorpc}: First,  it would greatly reduce our lack of
knowledge on nuclear parton densities, that for some observables causes
uncertainties of the same size as the eventual signal of high temperature,
deconfinement effects. Second, it would establish the QCD dynamics of relevance
for the LHC kinematic domain, were it the traditional fixed-order perturbative
one or a new non-linear perturbative regime. Finally, it would provide
theoretically based initial conditions for the macroscopic evolution of the system
and contribute to understanding the emergence of the hydrodynamical macroscopic
behaviour from the QCD microscopic dynamics.
\section{Concluding Remarks}
The future of deep inelastic scattering at high energies can be bright. The 
opportunity for founding and opening an new laboratory for energy frontier deep inelastic scattering 
which the LHC provides is indeed unique. It should not be disregarded  nor the huge LHC investment be thrown away. 
The LHeC is a one in a few decades  opportunity. 
Its science programme is unique and rich in its five components 
sketched here which all are dynamically advancing as a new generation
of theorists appears and the LHC physics becomes clearer. 
The huge step in energy and luminosity, which the LHeC represents, will almost certainly 
 entail surprises, unknown unknowns.
It is as well an important step to the development and exploitation of modern, low power, superconducting
 accelerator technology.
The detector represents a challenge many may wish to take up as the design of the 
HL LHC detector upgrades passes.
The LHeC  is far more than a mere step for the future of DIS because $ep$ and $eA$ scattering
at TeV energies are recognised to be intimate parts of modern particle 
and nuclear physics in general. 
This is underpinned by the observation and design that the $ep$ and $pp$ collisions
at the LHC may be recorded synchronously.

The next generation of machines at the energy
 frontier should comprise the LHeC for it is the only realistic option for energy frontier 
 $eh$ in the coming decades.  Not only
that the structure of matter is explored deeper, also new particles may be discovered
and new dynamics be found. The key carriers of hope and study of modern
collider physics, the $W,~Z$ bosons, the
top quark and the Higgs ought to and can  be investigated to sub-percent precision in $ep$ too.
A future big step in $pp$ collider energy requires a big step deeper into the dynamics inside the proton. 

Particle physics in order to proceed must utilise its rare opportunities,
the LHeC is a particularly attractive one, also for CERN's future. It can
be realised prior to the next, higher energy $pp$ collider, most likely the HE LHC,
which requires more time, more technology developments and substantially 
larger funds than the LHeC needs.  As Guido had put it {\it the physics case of the 
LHeC is made, can we proceed to build it}~\cite{guidoiac}. So sadly, this was the last word
which many of us heard him saying, in his unforgettable style.
It is time, $50$ years after the discovery of quarks at Stanford, to open a new
chapter for high energy electron-hadron scattering for which the LHC is the single
and luminous base. The linac to be built is shorter than the one Panofsky and colleagues mastered
to build at SLAC.\\

\noindent
$\bf{Acknowledgement}$ The development of the LHeC proceeds because it has been supported
by so many exceptional young and elder colleagues whom I just cannot mention here. I want to
thank Oliver Br\"uning for a decade of pleasant and fruitful co-coordination,
the CERN directorates under Rolf Heuer and Fabiola Gianotti for their supportive attitude,
the various active convenors on LHeC subjects, the FCC coordinators Michael Benedikt and Frank Zimmermann,
and the members of the International Advisory Committee led by  Herwig Schopper for their
encouraging and critical evaluation of the progress. The recent move towards the PERLE facility
at Orsay would not have been possible without the conviction of Achille Stocchi, shared with many
of us and not least the admirable teams of accelerator scientists  at CERN, Jlab and other institutes.
 For this article I enjoyed special help
by Claire Gwenlan, Uta Klein, Johannes Bl\"umlein, Peter Kostka, Frank Marhauser,  Oliver Fischer and Nestor Armesto. \\

\newpage 

\section*{Appendix 1: The Strong Coupling Constant}
\label{appalfas}
There was hardly a talk on deep inelastic scattering
 where Guido Altarelli would not speak about the importance and difficulty to determine the strong coupling constant. It is indeed amazing to see that we have been working on the determination of $\alpha_s$, or $\Lambda_{\overline{MS}}$, since the times
of the CDHS and BCDMS and other lepton-hadron scattering experiments, while 
others worked on $e^+e^-$ or $pp$, that is since 40 years, and yet we are not satisfied with the net 
result.
 In Guido's words: 
{\it{The crucial ingredient of perturbative QCD is the ``running" coupling  $\alpha_s(Q^2)$...
Some particularly simple and clear
hard processes must be used to measure the running coupling at a suitable energy scale .. The cleanest processes for measuring
$\alpha_s$ are the totally inclusive ones (no hadronic corrections) with light cone dominance,
 like $Z$ decay, scaling violations in DIS and perhaps $\tau$ decay (but, for $\tau$ , 
 the energy scale is dangerously low)...
In principle DIS is expected to be an ideal laboratory for the determination of 
$\alpha_s$ but in practice
the outcome is still to some extent open}}~\cite{Altarelli:2013bpa}. 

\subsection*{Status}
The coupling constant of the strong interaction, $\alpha_s$, is much less well known
than the fine-structure constant $\alpha$ and the Fermi constant $G_F$.  Its accurate and more precise 
determination is a key challenge for experimental and theoretical particle physics, and 
it is of fundamental importance. This has a number of reasons: 
\begin{itemize}
\item Approximately, within certain supersymmetric theories, the three couplings approach 
a point of convergence at the Planck scale. It is the large uncertainty of $\alpha_s$ which 
prevents strong conclusions to be drawn about this unification point as is illustrated in 
Fig.\,\ref{figalfgut}. This sets a goal to reach a ten times higher accuracy with the LHeC, see below.
\begin{figure}[h]
\vspace*{0.1cm}
\centerline{\includegraphics*[width=65mm,angle=-90.]{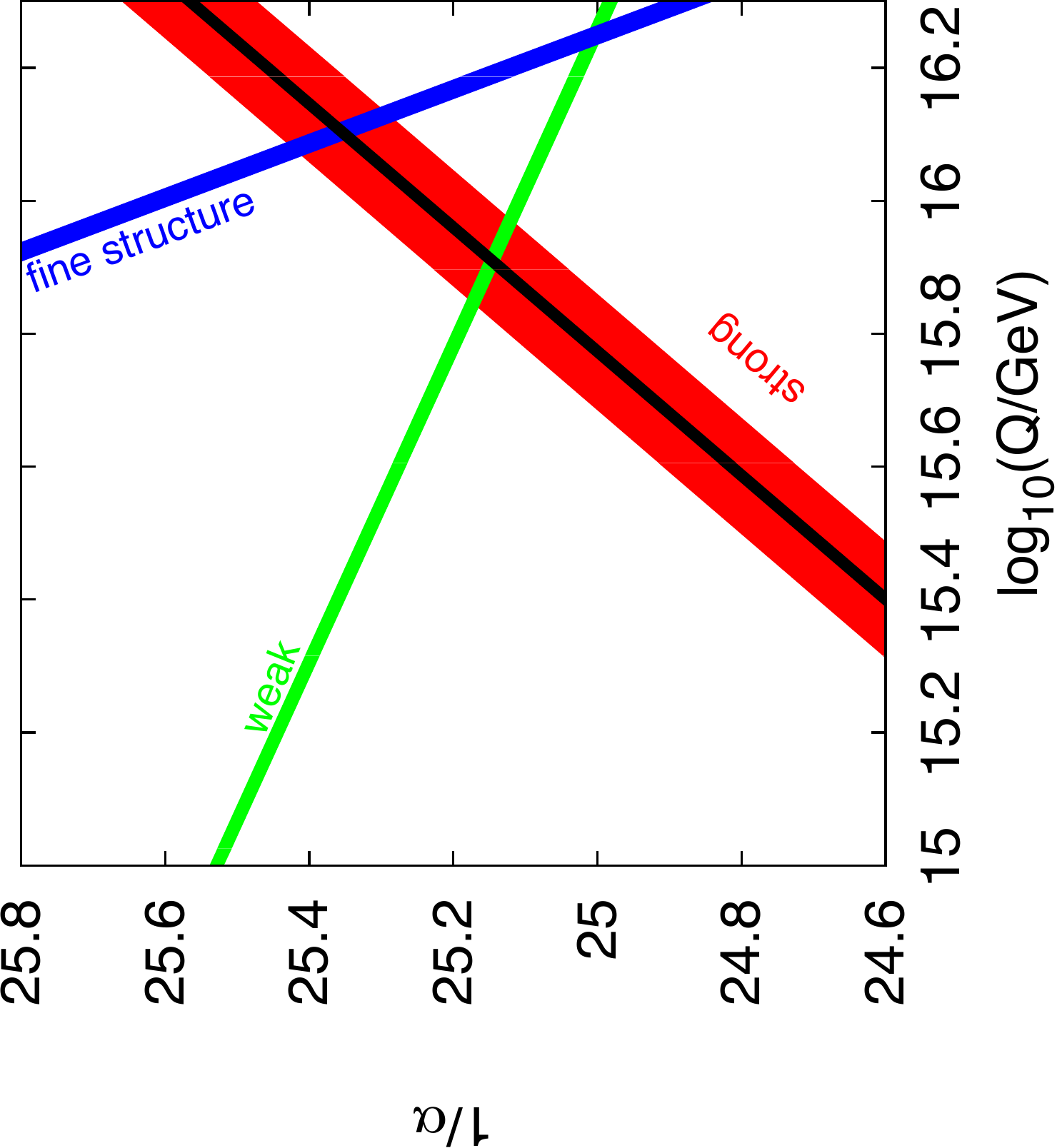}}
\caption{\footnotesize{Extrapolation of the inverse coupling constants to the Planck scale, illustrated
with a version of MMSM, from~\cite{AbelleiraFernandez:2012ty}. The coloured bands represent 
uncertainties as quoted by PDG in 2012, while the black insert to the red band 
on $\alpha_s$ illustrates the projected LHeC uncertainty, solely from inclusive DIS. The uncertainty of $\alpha_s$ 
quoted by PDG has recently increased by about $50$\,\%.}}
\label{figalfgut}
\end{figure}
\item The current average of various, different determinations of $\alpha_s$ is not satisfying which is discussed below and indicated by Guido's remark above. A single, better two independent, consistent measurements of $\alpha_s$ at per mille precision are the obvious, though difficult task ahead. In particular,
future $ep$ and $e^+e^-$ collider experiments must challenge the lattice
results, which also will improve.
\item Deep inelastic scattering has a special role in these determinations. The theory of inclusive DIS is clean, free of hadronic final state
phenomenology. In the past it led to peculiarities, due to lack of 
calibration redundance (BCDMS), smallness of $Q^2$ which  caused larger power (higher twist) corrections and nuclear and deuteron binding corrections. The most precise DIS data for $\alpha_s$, from BCDMS, point to a too low value. For long  a tendency had  been observed for jet analyses to provide larger values on $\alpha_s$ than inclusive ones.  All this will be overcome with the LHeC. At that time, DIS theory will have been calculated to N$^3$LO, and a new level of testing QCD and its development may be near.
\item With the LHC reaching high precision in various areas, the uncertainty of $\alpha_s$ becomes an
obstacle. This holds in particular for the derivation of Higgs couplings at the LHC, because the 
dominant production process is through gluon-gluon fusion, that is the product of $\alpha_s \cdot xg$
enters squared. An uncertainty of $\alpha_s(M_Z^2)$ taken to be $\pm 0.0015$, or $\pm 1.3$\,\%, translates to a $2.6$\,\% uncertainty on the cross section~\footnote{This is consistent with the result presented 
in~\cite{Bruening:2013bga}: using the iHixs code (v1.3)  the Higgs prodction cross section in 
$14$\,TeV $pp$ Drell-Yan scattering was calculated for NNPDF2.1 using two values of $\alpha_s$ which differed by $1.6$\,\%. The cross section for $0.121$ is larger than the one for $0.119$ by about $3.6$\,\%. One observes, as expected, that $\delta \sigma / \sigma \simeq 2
\delta \alpha_s / \alpha_s$.} 
 as obtained in~\cite{Anastasiou:2016cez}. This assumed uncertainty does not account for the different result obtained by the ABM team, see below. One clearly wishes to determine PDFs and
the strong coupling in a process suited to perform a genuine N$^3$LO pQCD analysis at a precision 
which should render effects from $\alpha_s$ and the parton distributions, discussed in Sect.\,\ref{secpdf}, negligible.
For Higgs physics at the LHC, this would be achieved with a few
per mille accuracy. Similar conclusions one derives when studying 
the PDF-$\alpha_s$ sensitivity of a high precision measurement of the $W$ boson mass~\cite{Aaboud:2017svj} or of the weak mixing angle at the LHC.
\end{itemize}

There exists a vast literature on $\alpha_s$ due to its fundamental importance, 
see e.g.~\cite{Deur:2016tte,Dissertori:2015tfa,dEnterria:2015kmd}.
The current status
of the determination of $\alpha_s$ has recently been summarised~\cite{Bethke:2017uli} and is 
reproduced in Tab.\,2 using the values as quoted in the PDG2016~\cite{Patrignani:2016xqp}.
\begin{table}[h]
\begin{center}
\begin{tabular}{|c|c|}
\hline
Method & $\alpha_s(M_Z^2)$ \\
\hline
Lattice QCD & $ 0.1184 \pm 0.0012$ \\
$\tau$-decays & $ 0.1192 \pm 0.0018$ \\
DIS & $ 0.1156 \pm 0.0021$ \\
Hadron Collider & $ 0.1151 \pm 0.0028$ \\
Electroweak Fits & $ 0.1196 \pm 0.0030$ \\
$e^+e^-$  & $ 0.1169 \pm 0.0034$ \\
\hline
\end{tabular}
\caption{\footnotesize{Mean values of the strong coupling 
constant $\alpha_s(M_Z^2)$, to NNLO and for $5$ active flavours~\cite{Patrignani:2016xqp}.}}
\end{center}
\label{tabaas}
\end{table}
A mean value has been derived of $\alpha_s(M_Z^2) = 0.1181 \pm 0.0011$ that  obviously is dominated
by the result of lattice calculations. These have recently been discussed in quite some 
detail~\cite{Aoki:2016frl}. In the lattice calculations the role of a measured cross section is taken 
by suitably defined Euclidean short distance quantities. Lattice calculations have a number
of additional, common peculiarities, they need input of the experimental hadronic spectrum and quark masses, they treat only light quarks with perturbative, matching additions of charm and beauty quark effects and they have uncertainties from
discretization and truncation of perturbative theory. There follows quite a range in the resulting $\alpha_s$ values obtained, beyond the simple value of uncertainty quoted, which is achieved by implementing certain quality criteria of the theoretical treatments as are presented in~\cite{Aoki:2016frl}. From a personal point of view one wished to keep the lattice results separate as PDG also does in providing an average  of
\begin{equation}
\alpha_s(M_Z^2) = 0.1174 \pm 0.0016
\end{equation}
for all  non-lattice results contained in Tab.\,2. The next accurate result is that from $\tau$-decays, which, however, as remarked by Guido above,  suffers from the small scale, i.e. largeness of $\alpha_s(\mu)$ 
at $\mu=M_{\tau}$ such that fixed order or contour improved perturbation theory lead to $7$\,\% 
different results, at $M_{\tau}$, and non-perturbative contributions to similarly sizeable 
uncertainties~\cite{Patrignani:2016xqp}. The next precise result is from deep inelastic scattering which
 has a number of partially well known peculiarities the LHeC will eventually resolve. 

\subsection*{Deep Inelastic Scattering on {\bf{$\alpha_s$}} - Now and with the LHeC}
\begin{figure}[h]
\centerline{
 \includegraphics[width=80mm]{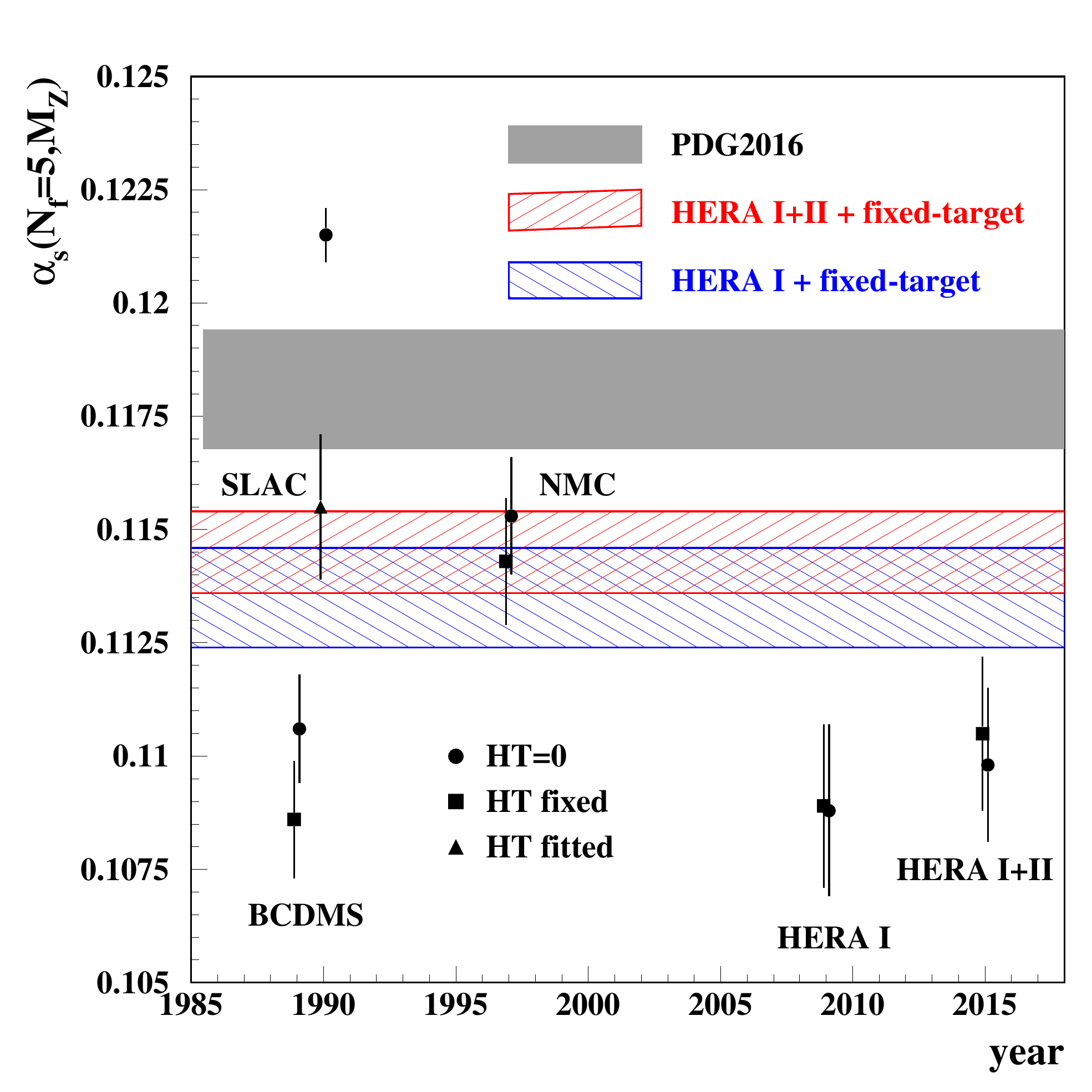}}
 \caption{\footnotesize{
   \label{figadis}
   Determinations of $\alpha_s$ by the ABMP group~\cite{Alekhin:2017kpj}.
   One observes quite some spread and a final result (red dashed) below
   the PDG average value, which includes DIS. Results are shown
   with the higher twist terms set to zero (circles),
   HT terms fixed to the values obtained in the ABMP16
   fit from considering all data sets (squares) and when fitting the 
    HT terms to the individual data sets (triangles).
   The bands  are obtained by using the combination  of the SLAC, BCDMS and NMC samples together with the combined H1 and ZEUS data from HERA I (left-tilted hatches, blue) and from HERA I+II (right-tilted hatches, red).  }}
\end{figure}

The status of the determination of $\alpha_s$ in DIS is well summarised in 
Fig.\,\ref{figadis} taken from a recent comprehensive  
paper~\cite{Alekhin:2017kpj} by the ABMP team. A big discrepancy is observed,
since long, of the large $\alpha_s \simeq 0.121$ value obtained 
with the SLAC data, at low $Q^2$ and large $x$, and the BCDMS result 
of $\alpha_s \simeq 0.11$. The cure to that observation had often been, 
as in~\cite{Alekhin:2017kpj} too,  the introduction of a higher twist correction
for $F_2$ which, as a power term $\propto Q^{-4}$, obviously only affects the 
SLAC result beyond the uncertainty, see Fig.\,\ref{figadis}. The fit to all DIS data by ABMP gives an NNLO
average of 
\begin{equation}
\alpha_s = 0.1140 \pm 0.0009
\end{equation}
 employing HT corrections and for $N_f=5$, which is drawn as the red shaded band for the ``HERA+fixed target data"-fit. The ABM result thus 
is (again) lower than the PDG average but consistent with the DIS value quoted from the PDG in Tab.\,2.
The overall NNLO value of $\alpha_s$ the ABMP group obtains, including also
hadron collider data, is $0.1147 \pm 0.0008$.  This may be compared, 
as was done in~\cite{Alekhin:2017kpj}, with the MSTW value of 
$0.1172 \pm 0.0013$ or NNPDF of $0.1174 \pm 0.0007$.

There have been quite some disputes over time about such a low value, the results of which Guido, above, called $to~some~extent~still~open$. 
It is in my view remarkable that an in depth analysis of the fixed target 
DIS (SLAC, NMC, BCDMS) and then available data by H1 came to similar conclusions as ABMP recently. 
The H1 Collaboration developed a dedicated procedure \cite{Adloff:2000qk}
for the determination of $\alpha_s$ which reduced the number of
parton distributions to just two for quarks, owing to the observation that $F_2$ 
in the proton  has a singlet, $\Sigma$, and a non-singlet, $\Delta$, component,
\begin{equation} \label{f2h1deco}
F_2 = \frac{2}{9} \cdot x \Sigma + \frac{1}{3} \cdot x \Delta = 
\frac{1}{3} \cdot xV + \frac{11}{9} \cdot xA.
\end{equation}
This introduced two functions $V$ and $A$, which approximately are valence 
and sea dominated, respectively. One therefore hoped to be able to reduce the 
uncertainty on $\alpha_s$ by reducing the PDF correlations to a minimum.
Similar observations to Fig.\,\ref{figadis} were made and in much detail
described in~\cite{Wallny:2001az}. The cure, however, was chosen to be different: because 
an unpleasant $Q^2_{min}$ dependence of $\alpha_s$ was observed, and to
avoid HT terms,
a cut of $Q^2 \geq 7.5$\,GeV$^2$ was implemented for the fixed target data. That was 
built into the bulk BCDMS data anyhow, as the detector had originally left space for a calorimeter to detect
CC scattering which was never put in place. It was then seen that the NMC data
had no further effect on $\alpha_s$,
due to the more precise BCDMS data. Moreover, inconsistencies
were observed and studied at length, see~\cite{Wallny:2001az},  of the  
slopes in $Q^2$ at high $x$ measured by SLAC and BCDMS. This caused the 
introduction of a cut for the inelasticity $y_{BCDMS}$ to be larger than $0.3$.
The result was that of full consistency, in the so constrained phase space, 
such that a $\chi^2 + 1$ criterion was applicable which rarely is the case
in global PDF and $\alpha_s$ analyses using unpleasant
 ``tolerance criteria" to screen the incompatibilities of data used. In 2001, H1 obtained an NLO 
$\alpha_s$ value of
\begin{equation} 
\label{alfash1}
\alpha_s(M_Z^2) = 0.1150~~\pm~~0.0017~(exp)~~^{+~~0.0009}_{-~~0.0005}~(model),
\end{equation}
which to NNLO pQCD would provide a lower number, probably close to $0.114$ as ABMP indeed now obtained.  
Here, the first error represents the experimental uncertainty on $\alpha_s$ 
of $1.5$\,\%.
 The second, lower
error includes all uncertainties associated with the construction of
the QCD model for the measured cross section. A
number to be noted may be that a $100$\,MeV uncertainty on the value of the charm
mass results in an $\alpha_s$ uncertainty of $\pm 0.0005$. A program to
determine $\alpha_s$ to per mille precision thus requires to fix the value of
$M_c$ to better than $10$\,MeV.

 When we on H1 designed our QCD fit
strategy, was it in 1990, I recall that the admirable Willy van Neerven advised to keep 
inclusive and jet data QCD analyses separate. It could be, he argued convincingly, 
that the gluon determined in inclusive DIS told us something
when compared to the jet data result, which obviously concerned $\alpha_s$ as well. 
It was therefore, for very long, avoided to put all together. In 2017, H1 published a 
comprehensive NNLO analysis of jet and all the
inclusive data~\cite{Andreev:2017vxu}. It led to the interesting
result, for jets only and to NNLO, of 
\begin{equation} 
\label{alfash1}
\alpha_s(M_Z^2) = 0.1157~~\pm~~0.0020~(exp)~~\pm~~0.0029~(thy).
\end{equation}
A joint fit to jet and inclusive DIS H1 data led to $\alpha_s=0.1142 \pm 0.0028$\, (tot), 
in remarkable agreement with ABMP. There again is a tendency to be noted that $\alpha_s$ 
from jets is somewhat higher than from inclusive DIS. This is about as far as HERA will 
lead and so we studied comprehensively the potential to obtain $\alpha_s$ from simulated LHeC data.

Two independent simulations and fit approaches have been undertaken
in order to verify the potential of the LHeC to determine $\alpha_s$,
see the CDR~\cite{Abelleira2012} for details. 
Table~\ref{tab:alfa} summarises the main results. It can be seen 
that the total experimental uncertainty on $\alpha_s$ is $0.2$\,\%
from the LHeC and $0.1$\,\% when combined with HERA. This determination
is free of higher twist, hadronic and nuclear corrections relying 
solely on inclusive DIS $ep$ data at high $Q^2$. 
\begin{table}
\begin{center}
\begin{tabular}{|l|c|c|c|}
\hline
case & cut [$Q^2$ (GeV$^2$)] & uncertainty & relative precision (\%)\\
\hline
HERA only & $Q^2>3.5$ & 0.00224 &  1.94  \\
HERA+jets & $Q^2>3.5$  & 0.00099 & 0.82 \\
\hline
LHeC only & $Q^2>3.5$  & 0.00020   & 0.17 \\
LHeC+HERA & $Q^2>3.5$  & 0.00013 & 0.11 \\
LHeC+HERA  & $Q^2>7.0$   & 0.00024   & 0.20 \\
LHeC+HERA  & $Q^2>10.$  & 0.00030 & 0.26  \\
\hline
\end{tabular}
\caption{\footnotesize{Results of NLO QCD fits to HERA data (top, without and with jets)
to the simulated LHeC data alone and to their combination,
for details of the fit see ~\cite{Abelleira2012}.
The resulting uncertainty includes all the statistical and experimental
systematic error sources taking their correlations into account. The LHeC
result does not include jet data.
}}
\label{tab:alfa}
\end{center}
\end{table}
There are known further parametric uncertainties in DIS determinations
of $\alpha_s$. These can safely be expected to be much reduced
also by the LHeC, which promises to determine the charm mass, for
example, by $3$\,MeV, as compared to $40$\,MeV at HERA, corresponding
to an $\alpha_s$ uncertainty of $0.04$\,\%. Matching the
experimental uncertainty requires
that when the LHeC operates such analyses were performed in N$^3$LO
pQCD in order to reduce the renormalisation and factorisation
scale uncertainty. 
Due to the huge range
in $Q^2$ and the high precision of the data new, decisive tests will
also become available for answering the question whether the
strong coupling determined with jets and in inclusive DIS
are the same. If confirmed, as is demonstrated in Tab.\,\ref{tab:alfa}
with the HERA data, a joint inclusive and jet analysis has the 
potential to even further reduce the uncertainty of $\alpha_s$
derived from future LHeC data.
The ambition to
measure $\alpha_s$ to per mille accuracy therefore represents
a vision for a renaissance of the physics of deep inelastic scattering
which is a major goal of the whole LHeC enterprise.

\section*{Appendix 2: The Longitudinal Structure Function and R}
In 2012, Guido Altarelli publishes an article~\cite{Altarelli:2011zv}, in honour 
of Mario Greco, about ``The Early Days of QCD (as seen from Rome)'' 
in which he describes one of his main achievements, and long term irritation, regarding 
the so-called longitudinal structure function, $F_L$,  and its measurement.
 {\it{
Back to Rome I met Guido Martinelli, then a post-doc with a contract for doing
accelerator physics at Frascati, and I rescued him into particle physics, with a work on
the transverse momentum distributions for jets in lepto-production final 
states~\cite{Altarelli:1978tq}. In
the same paper we derived an elegant formula for the longitudinal structure function
$F_L$, also an effect of order $\alpha_s(Q^2)$, as a convolution integral over $F_2(x,Q^2)$ and the
gluon density $g(x,Q^2)$. I find it surprising that it took 40 years since the start of deep
inelastic scattering experiments to get meaningful data on the longitudinal structure
function. The present data, recently obtained by the H1 experiment at DESY, are in
agreement with this LO QCD prediction but the accuracy of the test is still far from
being satisfactory for such a basic quantity.}} 

The measurement of $F_L$ concerned me for more than twenty years, from first 
studies of the sensitivity at HERA in 1988 to the
first announcement of results at in Philadelphia ICHEP 2008~\cite{Klein:2008xz}
and its final publication. Why 
is $F_L$ a quantity of special interest, and why was its determination so difficult at 
HERA and how would LHeC surpass it?

The inclusive, deep inelastic electron-proton scattering cross section at low $Q^2$,
written in its reduced form,
\begin{equation}
 \sigma_r = \frac{d^2\sigma}{dxdQ^2} \cdot    \frac{Q^4 x} {2\pi \alpha^2 Y_+} =  
    F_2(x,Q^2) - \frac{y^2}{Y_+} \cdot F_L(x,Q^2)
       \label{sig}
  \end{equation}  
is defined by two proton structure functions, $F_2$ and $F_L$, where
$Q^2$ is the negative four-momentum transfer squared, $y$ the inelasticity
$y=Q^2/sx \leq 1$, $Y_+ = 1+ (1-y)^2$ and $\alpha$ is
the fine structure constant. The cross section may also be 
expressed~\cite{Hand:1963zz} as a 
sum of two contributions, $\sigma_r \propto (\sigma_T + \epsilon \sigma_L)$,
referring to the transverse and longitudinal polarisation state of the exchanged boson, 
with $\epsilon$ characterising the ratio of the longitudinal to the transverse 
polarisation. The ratio of the longitudinal to transverse cross sections is termed
\begin{equation}
R(x,Q^2) = \frac{\sigma_L}{\sigma_T} = \frac{F_L}{F_2-F_L},
\label{R}
\end{equation}
which is related to $F_2$ and $F_L$ as given above. Due to the positivity of the 
cross sections $\sigma_{L,T}$ one observes that $F_L \leq F_2$. The
reduced cross section $\sigma_r$, for $Q^2$ much below the weak boson masses, 
is therefore a direct measure of $F_2$. That holds apart from a limited region of high $y$
where a contribution of $F_L$ may be sizeable. The principal task to measure $F_L$
thus requires to precisely measure the inclusive DIS cross section near to $y=1$, and
then to disentangle the two structure functions by utilising the $y^2/Y_+$ variation
which depends on $x$, $Q^2$ and $s$, that is one may vary $s$ and obtains $F_2$
and $F_L$ at a common, fixed point of $x,Q^2$. That is challenging not only because  
the $F_L$ part is small but because $y = 1 - E'/E_e$ is large only for scattered electron
energies $E'$ much smaller than the electron beam energy $E_e$, e.g. $E' = 2.7$\,GeV 
for $y=0.9$ at HERA. In the region where $E'$ is a few GeV only, the electron identification 
becomes a major problem and the electromagnetic ($\pi^0 \rightarrow \gamma \gamma$) 
and hadronic background, mainly from unrecognised photo-production, rises enormously.

The interest in a measurement of $F_L$, especially at low $x$, is related to the 
uncertainty in the determination of the gluon distribution, $xg(x,Q^2)$, in pQCD
fit extractions of the quark and gluon densities, besides $\alpha_s$. On more
theoretical grounds, there is an expectation 
that at small $x$ the linear DGLAP $Q^2$ evolution equations, which 
give $xg$ a certain meaning, may eventually have to be replaced by a BFKL type
of law. The gluon density is defined in a theoretical framework which needs to
be confirmed or disproven which on the basis of only $F_2$ is just not possible. 
With $F_2$ and $F_L$, however, one has two principal handles for an extraction of
$xg$ and the main motivation for our proposal~\cite{H1FL05} to operate HERA at lower energies
was to test higher order QCD at small $x$.

The theoretically most sound and experimentally most precise 
determination of the gluon density $xg$ is through
the measurement of the $Q^2$ dependence of $F_2$ in 
DIS\footnote{This is transparent from the sudden increase of uncertainty
for $xg$ below $x \simeq 0.0005$, see Fig.\,\ref{figxg}, where HERA's access to
$\partial F_{2}/ \partial \ln Q^{2}$ ends.}. Their relation is prescribed 
by the DGLAP equations
\begin{equation}
        \frac{\partial F_{2}(x,Q^2)}{  \partial \ln Q^{2}} = \frac{\alpha_s(Q^2)}{2 \pi} 
        \int_x^1 dz  \left[ F_2(\frac{x}{z}) P_{qq}(z) + 2 
        \sum_{i=1}^{N_f} e_i^2 \cdot G(\frac{x}{z}) P_{qG}(z) \right]
\label{dglap}
\end{equation}
with the splitting functions $P$, $xg=G$, $e_i$ the electric charge of quark $i$, 
and $N_f$ the number of active flavours. At small $x$ the first term is almost negligible. 
One thus finds that the measurement of the
$\ln Q^2$ derivative of $F_2$ determines $xg$, i.e. one needs very high precision in $F_2(x,Q^2)$ 
combined with a large lever arm in $Q^2$ at each value of $x$. 
The extraction of the $F_2$ derivative had 
been evaluated in quite some detail by the H1 Collaboration when 
first precise measurements 
on $F_2$ became available, see~\cite{Adloff:2000qk}. It was also 
instrumental to the discovery~\cite{Abt:1993kh} of the rise of $xg$ towards low $x$. 

The second method to cleanly access $xg$ is through the relation mentioned by Guido
\begin{equation}
        F_L(x,Q^2) = \frac{\alpha_s}{\pi} x^2
        \int_x^1 \frac{dz}{z^3}  \left[ \frac{4}{3}F_2(z,Q^2) + 
       2 \sum_i^{N_f} e_i^2 \cdot G(z,Q^2) \left(1-\frac{x}{z} \right)  \right]
\label{altmar}
\end{equation}
derived in~\cite{Altarelli:1978tq}, see also~\cite{Gluck:1980cp}.
One can calculate that the integrand at small $x$
 is dominated by the gluon distribution term which 
directly relates $F_L$ and $xg$.

\begin{figure}[h]
\centerline{\includegraphics*[width=75mm]{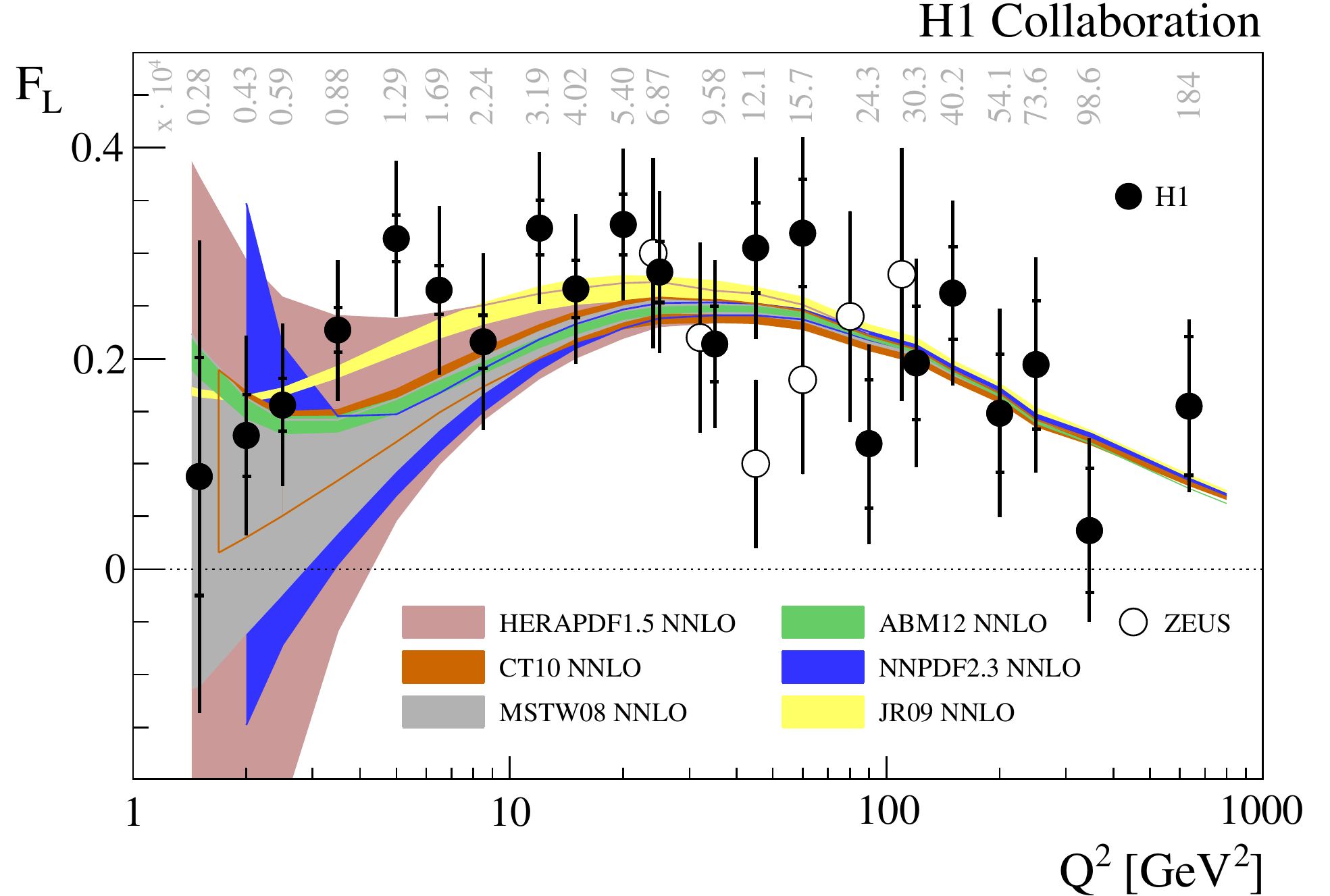}
\includegraphics*[width=75mm]{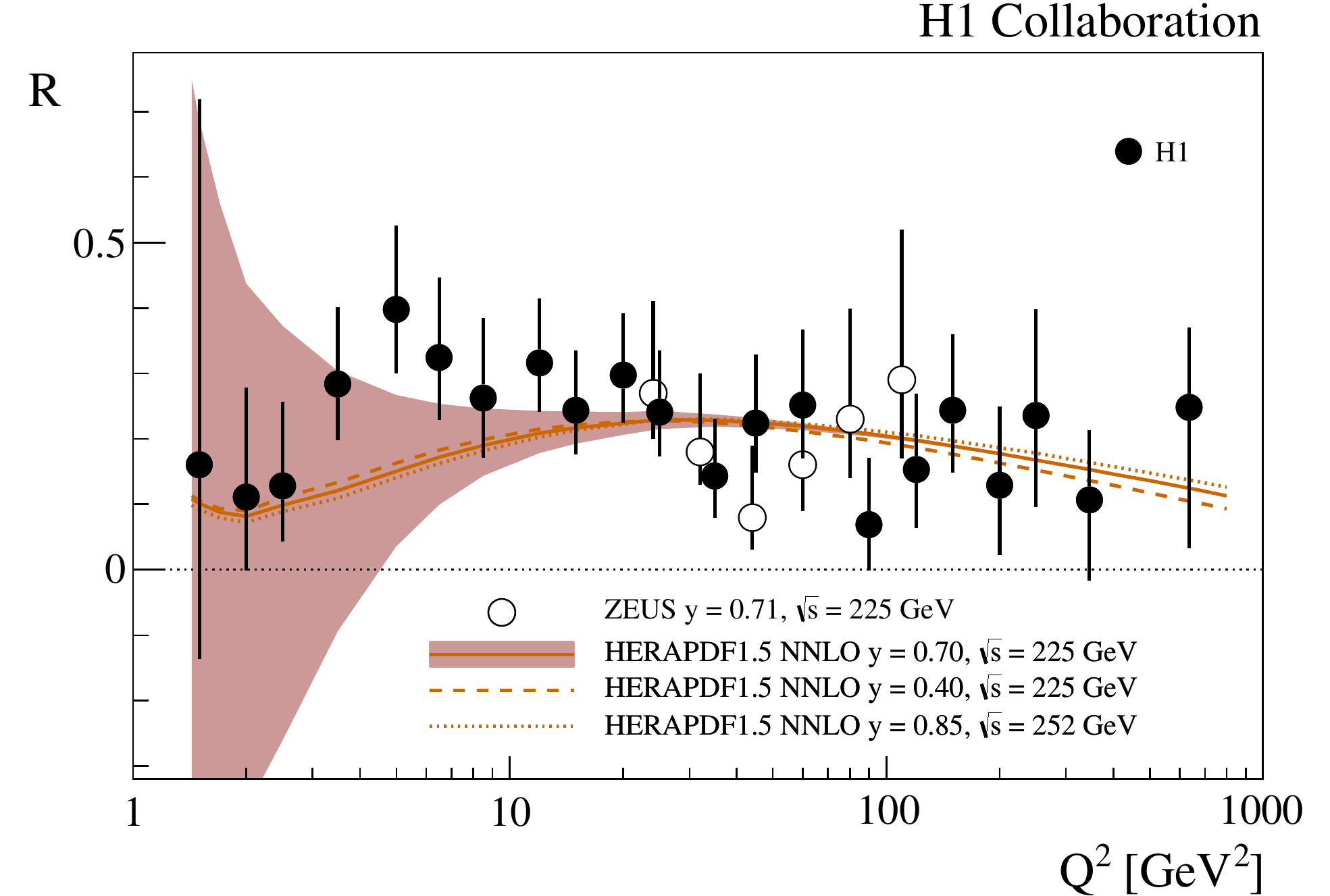}}
\caption{\footnotesize{Left: Measurement of the structure function $F_L(x,Q^2)$
at H1 (solid points) and ZEUS (open circles), from a variation of 
proton beam energy in the final half year of HERA operation.
The curves and their uncertainty bands illustrate predictions on $F_L$ obtained
to NNLO pQCD by various groups engaged in the extraction of PDFs.
Right: Measurement of the structure function ratio $R=F_L/(F_2-F_L)$
at H1 (solid points) and ZEUS (open circles), from a variation of 
proton beam energy in the final half year of HERA operation. The curve represents 
an NNLO QCD fit analysis of the other HERA data. This becomes uncertain for $Q^2$ below
$10$\,GeV$^2$ where the $Q^2$ dependence of $F_2$ does not permit
an accurate determination of the gluon density which dominates the prediction on $F_L$, see text.
}}
\label{figflh1}
\end{figure}

The final measurement of $F_L$ at HERA~\cite{Andreev:2013vha}
is illustrated in Fig.\,\ref{figflh1}.
The data were obtained in an evolved separation procedure which principally
determined  $F_L$ as the intercept, at $y=0$, and $F_L$
as the negative slope of  the linear dependence of  $\sigma_r$ 
on $y^2/Y_+$.  The variation
in $y$ was achieved at HERA by comparing high  statistics data 
at highest energy, $E_p=920$\,GeV, with about $13$\,pb$^{-1}$ of  data at $460$\,GeV
and $7$\,pb$^{-1}$ at $575$\,GeV. The values of $E_p$ had been
chosen for about equidistant separation of the measurements in
$y^2/Y_+$. The low energy runs took place from March to 
June 2007, with only a few days of setup time for the machine.
One notices that the absolute, total uncertainty of the H1 and ZEUS data
is about $\pm 0.1$, with notable statistical uncertainties for $Q^2 \geq 20$\,GeV$^2$.
The ZEUS data do not extend to low $Q^2$ values, which  could be accessed
by H1~\cite{Collaboration:2010ry} thanks to the existence of the Backward 
Silicon Tracker which my group had developed~\cite{Eick:1996gv}, built and operated over a time of 15 years. 
A clear track-cluster link was required to identify low energy scattered electrons
 near the beam pipe. The determination of the electron's charge, in a demanding, small radius
curvature measurement, permitted in addition the removal of
a large part of background with a residual account for charge asymmetry effects as
are caused by the difference in energy depositions of anti-protons and protons at low energy. 
As is described in~\cite{Collaboration:2010ry} quite refined techniques had to be 
found and to work for accessing $F_L$.

In Fig.\,\ref{figflh1} the data are compared with predictions, calculated in NNLO pQCD, from various groups who perform comprehensive analyses of parton distribution functions.
One observes a generally good agreement and recognises the origin for Guido
Altarelli's remark cited above: the data are impressive, however, they are not
precise enough for crucial tests of pQCD. In the specially interesting region of low $x$  where the PDF uncertainties, mainly related to $xg$ (see Sect.\,\ref{secpdf}), are substantial, the data hint to a positive and regular $F_L$ but lack discriminative
power also. This can also be illustrated well with the measurement of $\sigma_L/\sigma_T$.

The first measurements of $R$ at SLAC~\cite{Miller:1971qb,Riordan:1974te},
in the range of $Q^2$ between $1$ and 
about $10$\,GeV$^2$ and $0.1 \leq x \leq 0.5$,  found the ratio to be about $0.18$, i.e. small
enough to be taken as evidence for quarks to carry spin $1/2$. The HERA data
determine this ratio as a function of $Q^2$, and correspondingly 
varying $x =Q^2/(4 E_e E_p y) \simeq Q^2/ (4 \cdot 10^4 GeV^2)$, for $E_p=575$\,GeV and
$y \simeq 0.7$. The result is displayed in Fig.\,\ref{figflh1}.  To good approximation,
$R(x,Q^2)$ is a constant which was determined as $R=0.23 \pm 0.04$, in 
astonishing good agreement with the SLAC values. The resulting value also obeys
the condition, $R \leq 0.37$, that had been obtained in a rigorous attempt to derive the 
dipole model for inelastic DIS~\cite{Ewerz:2006vd}.

The LHeC will lead to the first precision measurement of $F_L$ and $R$. This is due to
three main improvements: i) the luminosity is expected to be so large that high statistics,
fb$^{-1}$ measurements can be performed at even more than two low beam energy points, compared with the two $10$\,pb$^{-1}$  sets at HERA; ii) the apparatus in the backward region is being designed as a high precision device with more Silicon detector planes of higher  acceptance and resolution and a hadronic backward calorimeter which was basically absent on H1; iii) the increased electron beam energy implies that high $y$ may be achieved at larger scattered electron energy $E'$. Both
the improved detector and the enlarged $E_e$ will enable to reach highest $y$ values at much reduced background. 

\begin{figure}[tb]
\centerline{\includegraphics*[width=100mm]{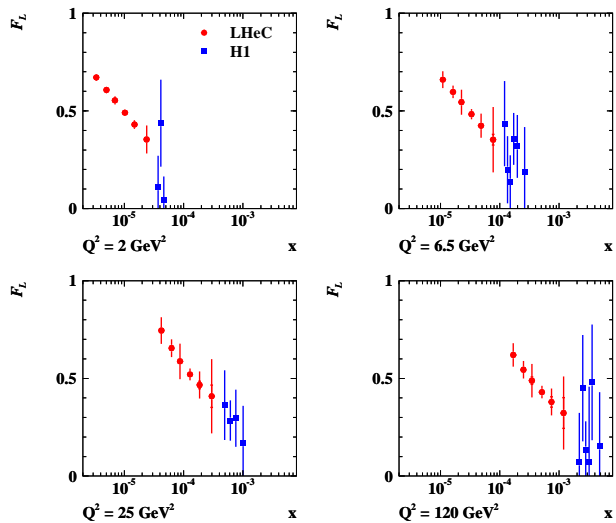}}
\caption{\footnotesize{Simulations of $F_L$ measurements with the LHeC (red circles) compared
with measurements at H1 (blue squares), see text.}}
\label{figflhec}
\end{figure}

A simulation had been performed for the LHeC CDR~\cite{Abelleira2012} which is 
illustrated in Fig.\,\ref{figflhec}. In order to be conceptually independent of the 
LHC operation, for the LHeC the electron beam energy is lowered as opposed to HERA.
The point-by-point precision is impressively improved, 
from at best $\delta F_L \simeq \pm 0.1-0.2$ with H1 to typically a $0.02$ total uncertainty for the LHeC. 
Based on the invaluable experience gained 
with H1 at HERA and on the design prospects for the LHeC and its $ep$ experiment, one 
can indeed be optimistic that Guido Altarelli's wish for a precise determination 
of $F_L$ will eventually be fulfilled. The simulated  data, with their exceptional determinations of $F_2$ and $F_L$,
were used in a study, presented in the CDR, to illustrate the unique potential in 
discriminating theory at small $x$.

%

%
+
\footnotesize{
\bibliography{lhecbibga.tex}
}
\end{document}